\input harvmac.tex
 \input epsf.tex
 \input amssym

\def\figin{\epsfcheck\figin}\def\figins{\epsfcheck\figins}
\def\epsfcheck{\ifx\epsfbox\UnDeFiNeD
\message{(NO epsf.tex, FIGURES WILL BE IGNORED)}
\gdef\figin##1{\vskip2in}\gdef\figins##1{\hskip.5in}
\else\message{(FIGURES WILL BE INCLUDED)}%
\gdef\figin##1{##1}\gdef\figins##1{##1}\fi}
\def\DefWarn#1{}
\def\figinsert{\goodbreak\midinsert}
\def\ifig#1#2#3{\DefWarn#1\xdef#1{fig.~\the\figno}
\writedef{#1\leftbracket fig.\noexpand~\the\figno} %
\figinsert\figin{\centerline{#3}}\medskip\centerline{\vbox{\baselineskip12pt
\advance\hsize by -1truein\noindent\footnotefont{\bf
Fig.~\the\figno:} #2}}
\bigskip\endinsert\global\advance\figno by1}



\def \beq  {\begin{eqnarray}}
\def \eeq  {\end{eqnarray}}


\def\frac#1#2{{#1 \over #2}}
\def\text#1{#1}
\def\bz{\bar{z}}


\lref\RyuBV{
  S.~Ryu and T.~Takayanagi,
  ``Holographic derivation of entanglement entropy from AdS/CFT,''
Phys.\ Rev.\ Lett.\  {\bf 96}, 181602 (2006).
[hep-th/0603001].
}

\lref\HubenyXT{
  V.~E.~Hubeny, M.~Rangamani and T.~Takayanagi,
  ``A Covariant holographic entanglement entropy proposal,''
JHEP {\bf 0707}, 062 (2007).
[arXiv:0705.0016 [hep-th]].
}

\lref\HeadrickKM{
  M.~Headrick and T.~Takayanagi,
  ``A Holographic proof of the strong subadditivity of entanglement entropy,''
Phys.\ Rev.\ D {\bf 76}, 106013 (2007).
[arXiv:0704.3719 [hep-th]].
}

\lref\HubenyRE{
  V.~E.~Hubeny and M.~Rangamani,
  ``Holographic entanglement entropy for disconnected regions,''
JHEP {\bf 0803}, 006 (2008).
[arXiv:0711.4118 [hep-th]].
}

\lref\AbajoArrastiaYT{
  J.~Abajo-Arrastia, J.~Aparicio and E.~Lopez,
  ``Holographic Evolution of Entanglement Entropy,''
JHEP {\bf 1011}, 149 (2010).
[arXiv:1006.4090 [hep-th]].
}

\lref\AparicioZY{
  J.~Aparicio and E.~Lopez,
  ``Evolution of Two-Point Functions from Holography,''
JHEP {\bf 1112}, 082 (2011).
[arXiv:1109.3571 [hep-th]].
}

\lref\OstlundZZ{
  S.~Ostlund and S.~Rommer,
  ``Thermodynamic Limit of Density Matrix Renormalization for the spin-1 heisenberg chain,''
Phys.\ Rev.\ Lett.\  {\bf 75}, 3537 (1995).
[cond-mat/9503107].
}

\lref\cirac{
F. Verstraete and J. I. Cirac,
``Renormalization algorithms for Quantum-Many Body Systems in two and higher dimensions,"
[arXiv:cond36
mat/0407066v1] (2004).
}

\lref\AlbashMV{
  T.~Albash and C.~V.~Johnson,
  ``Evolution of Holographic Entanglement Entropy after Thermal and Electromagnetic Quenches,''
New J.\ Phys.\  {\bf 13}, 045017 (2011).
[arXiv:1008.3027 [hep-th]].
}

\lref\TakayanagiWP{
  T.~Takayanagi and T.~Ugajin,
  ``Measuring Black Hole Formations by Entanglement Entropy via Coarse-Graining,''
JHEP {\bf 1011}, 054 (2010).
[arXiv:1008.3439 [hep-th]].
}

\lref\MaldacenaKR{
  J.~M.~Maldacena,
  ``Eternal black holes in anti-de Sitter,''
JHEP {\bf 0304}, 021 (2003).
[hep-th/0106112].
}

\lref\CalabreseEU{
  P.~Calabrese and J.~L.~Cardy,
  ``Entanglement entropy and quantum field theory,''
J.\ Stat.\ Mech.\  {\bf 0406}, P06002 (2004).
[hep-th/0405152].
}
\lref\IsraelUR{
  W.~Israel,
  ``Thermo field dynamics of black holes,''
Phys.\ Lett.\ A {\bf 57}, 107 (1976)..
}

\lref\NishiokaUN{
  T.~Nishioka, S.~Ryu and T.~Takayanagi,
  ``Holographic Entanglement Entropy: An Overview,''
J.\ Phys.\ A {\bf 42}, 504008 (2009).
[arXiv:0905.0932 [hep-th]].
}

\lref\CalabreseIN{
  P.~Calabrese and J.~L.~Cardy,
  ``Evolution of entanglement entropy in one-dimensional systems,''
J.\ Stat.\ Mech.\  {\bf 0504}, P04010 (2005).
[cond-mat/0503393].
}
\lref\CalabreseRG{
  P.~Calabrese and J.~Cardy,
  ``Quantum Quenches in Extended Systems,''
J.\ Stat.\ Mech.\  {\bf 0706}, P06008 (2007).
[arXiv:0704.1880 [cond-mat.stat-mech]].
}

\lref\CalabreseRX{
  P.~Calabrese and J.~L.~Cardy,
  ``Time-dependence of correlation functions following a quantum quench,''
Phys.\ Rev.\ Lett.\  {\bf 96}, 136801 (2006).
[cond-mat/0601225].
}

\lref\CalabreseQY{
  P.~Calabrese and J.~Cardy,
  ``Entanglement entropy and conformal field theory,''
J.\ Phys.\ A {\bf 42}, 504005 (2009).
[arXiv:0905.4013 [cond-mat.stat-mech]].
}

\lref\MorrisonIZ{
  I.~A.~Morrison and M.~M.~Roberts,
  ``Mutual information between thermo-field doubles and disconnected holographic boundaries,''
[arXiv:1211.2887 [hep-th]].
}

\lref\BanadosWN{
  M.~Banados, C.~Teitelboim and J.~Zanelli,
 ``The Black hole in three-dimensional space-time,''
Phys.\ Rev.\ Lett.\  {\bf 69}, 1849 (1992).
[hep-th/9204099].
}

\lref\MaldacenaBW{
  J.~M.~Maldacena and A.~Strominger,
  ``AdS(3) black holes and a stringy exclusion principle,''
JHEP {\bf 9812}, 005 (1998).
[hep-th/9804085].
}

\lref\HeadrickZT{
  M.~Headrick,
  ``Entanglement Renyi entropies in holographic theories,''
Phys.\ Rev.\ D {\bf 82}, 126010 (2010).
[arXiv:1006.0047 [hep-th]].
}

\lref\SiopsisUP{
  G.~Siopsis,
  ``Large mass expansion of quasinormal modes in AdS(5),''
Phys.\ Lett.\ B {\bf 590}, 105 (2004).
[hep-th/0402083].
}

\lref\LoukoHC{
  J.~Louko and D.~Marolf,
  ``Single exterior black holes and the AdS / CFT conjecture,''
Phys.\ Rev.\ D {\bf 59}, 066002 (1999).
[hep-th/9808081].
}

\lref\FestucciaZX{
  G.~Festuccia and H.~Liu,
  ``A Bohr-Sommerfeld quantization formula for quasinormal frequencies of AdS black holes,''
Adv.\ Sci.\ Lett.\  {\bf 2}, 221 (2009).
[arXiv:0811.1033 [gr-qc]].
}

\lref\MaldacenaXP{
  J.~Maldacena and G.~L.~Pimentel,
  ``Entanglement entropy in de Sitter space,''
JHEP {\bf 1302}, 038 (2013).
[arXiv:1210.7244 [hep-th]].
}

\lref\BriganteJV{
  M.~Brigante, G.~Festuccia and H.~Liu,
  ``Hagedorn divergences and tachyon potential,''
JHEP {\bf 0706}, 008 (2007).
[hep-th/0701205].
}

\lref\FestucciaSA{
  G.~Festuccia and H.~Liu,
  ``The Arrow of time, black holes, and quantum mixing of large N Yang-Mills theories,''
JHEP {\bf 0712}, 027 (2007).
[hep-th/0611098].
}

\lref\BriganteBQ{
  M.~Brigante, G.~Festuccia and H.~Liu,
  ``Inheritance principle and non-renormalization theorems at finite temperature,''
Phys.\ Lett.\ B {\bf 638}, 538 (2006).
[hep-th/0509117].
}

\lref\FestucciaPI{
  G.~Festuccia and H.~Liu,
  ``Excursions beyond the horizon: Black hole singularities in Yang-Mills theories. I.,''
JHEP {\bf 0604}, 044 (2006).
[hep-th/0506202].
}

\lref\FidkowskiNF{
  L.~Fidkowski, V.~Hubeny, M.~Kleban and S.~Shenker,
  ``The Black hole singularity in AdS / CFT,''
JHEP {\bf 0402}, 014 (2004).
[hep-th/0306170].
}

\lref\SwingleBG{
  B.~Swingle,
  ``Entanglement Renormalization and Holography,''
Phys.\ Rev.\ D {\bf 86}, 065007 (2012).
[arXiv:0905.1317 [cond-mat.str-el]].
}

\lref\BalasubramanianCE{
  V.~Balasubramanian, A.~Bernamonti, J.~de Boer, N.~Copland, B.~Craps, E.~Keski-Vakkuri, B.~Muller and A.~Schafer {\it et al.},
  ``Thermalization of Strongly Coupled Field Theories,''
Phys.\ Rev.\ Lett.\  {\bf 106}, 191601 (2011).
[arXiv:1012.4753 [hep-th]].
}

\lref\BalasubramanianUR{
  V.~Balasubramanian, A.~Bernamonti, J.~de Boer, N.~Copland, B.~Craps, E.~Keski-Vakkuri, B.~Muller and A.~Schafer {\it et al.},
  ``Holographic Thermalization,''
Phys.\ Rev.\ D {\bf 84}, 026010 (2011).
[arXiv:1103.2683 [hep-th]].
}

\lref\AsplundCQ{
  C.~T.~Asplund and S.~G.~Avery,
  ``Evolution of Entanglement Entropy in the D1-D5 Brane System,''
Phys.\ Rev.\ D {\bf 84}, 124053 (2011).
[arXiv:1108.2510 [hep-th]].
}

\lref\BasuFT{
  P.~Basu and S.~R.~Das,
  ``Quantum Quench across a Holographic Critical Point,''
JHEP {\bf 1201}, 103 (2012).
[arXiv:1109.3909 [hep-th]].
}

\lref\BasuGG{
  P.~Basu, D.~Das, S.~R.~Das and T.~Nishioka,
  ``Quantum Quench Across a Zero Temperature Holographic Superfluid Transition,''
[arXiv:1211.7076 [hep-th]].
}

\lref\BalasubramanianAT{
  V.~Balasubramanian, A.~Bernamonti, N.~Copland, B.~Craps and F.~Galli,
  ``Thermalization of mutual and tripartite information in strongly coupled two dimensional conformal field theories,''
Phys.\ Rev.\ D {\bf 84}, 105017 (2011).
[arXiv:1110.0488 [hep-th]].
}

\lref\AllaisYS{
  A.~Allais and E.~Tonni,
  ``Holographic evolution of the mutual information,''
JHEP {\bf 1201}, 102 (2012).
[arXiv:1110.1607 [hep-th]].
}

\lref\HubenyRY{
  V.~E.~Hubeny,
  ``Extremal surfaces as bulk probes in AdS/CFT,''
JHEP {\bf 1207}, 093 (2012).
[arXiv:1203.1044 [hep-th]].
}

\lref\MaldacenaXP{
  J.~Maldacena and G.~L.~Pimentel,
  ``Entanglement entropy in de Sitter space,''
JHEP {\bf 1302}, 038 (2013).
[arXiv:1210.7244 [hep-th]].
}

\lref\BuchelLLA{
  A.~Buchel, L.~Lehner, R.~C.~Myers and A.~van Niekerk,
  ``Quantum quenches of holographic plasmas,''
[arXiv:1302.2924 [hep-th]].
}
\lref\VidalMera{
G.~Vidal,
``A class of quantum many-body states that can be efficiently simulated,"
Phys.~Rev.~Lett.~101, 110501 (2008).
[arXiv: quant-ph/0610099].
}

\lref\TakayanagiZK{
  T.~Takayanagi,
  ``Holographic Dual of BCFT,''
Phys.\ Rev.\ Lett.\  {\bf 107}, 101602 (2011).
[arXiv:1105.5165 [hep-th]].
}

\lref\BoussoSJ{
  R.~Bousso, S.~Leichenauer and V.~Rosenhaus,
  ``Light-sheets and AdS/CFT,''
Phys.\ Rev.\ D {\bf 86}, 046009 (2012).
[arXiv:1203.6619 [hep-th]].
}

\lref\CzechBH{
  B.~Czech, J.~L.~Karczmarek, F.~Nogueira and M.~Van Raamsdonk,
  ``The Gravity Dual of a Density Matrix,''
Class.\ Quant.\ Grav.\  {\bf 29}, 155009 (2012).
[arXiv:1204.1330 [hep-th]].
}

\lref\BoussoMH{
  R.~Bousso, B.~Freivogel, S.~Leichenauer, V.~Rosenhaus and C.~Zukowski,
  ``Null Geodesics, Local CFT Operators and AdS/CFT for Subregions,''
[arXiv:1209.4641 [hep-th]].
}

\lref\VidalHolo{
 G.~Evenbly and G.~Vidal,
 ``Tensor network states and geometry,"
J.~Stat.~Phys.~{\bf 145}, 891 (2011).
[arXiv: 1106.1082 [quant-ph]].
}

\lref\VidalRev{
G.~Vidal,
``Entanglement Renormalization: an introduction,"
in {\it Understanding Quantum Phase Transitions}, L.~D.~Carr ed. (Taylor \& Francis, Boca Raton, 2010).
[arXiv: 0912.1651 [cond-mat.str-el]].
}

\lref\SchollRev{
U.~Schollwoeck,
``The density-matrix renormalization group in the age of matrix product states,"
Ann.~Phys.~{\bf 326}, 96 (2011).
[arXiv:1008.3477 [cond-mat.str-el]].
}

\lref\VidalTree{
Y.~Shi, L.~Duan, G.~Vidal,
``Classical simulation of quantum many-body systems with a tree tensor network,"
Phys.~Rev.~A {\bf 74}, 022320 (2006).
[arXiv: quant-ph/0511070].
}

\lref\SwingleWQ{
  B.~Swingle,
  ``Constructing holographic spacetimes using entanglement renormalization,''
[arXiv:1209.3304 [hep-th]].
}

\lref\NozakiZJ{
  M.~Nozaki, S.~Ryu and T.~Takayanagi,
  ``Holographic Geometry of Entanglement Renormalization in Quantum Field Theories,''
[arXiv:1208.3469 [hep-th]].
}

\lref\WhiteZZ{
  S.~R.~White,
  ``Density matrix formulation for quantum renormalization groups,''
Phys.\ Rev.\ Lett.\  {\bf 69}, 2863 (1992)..
}

\lref\HolzheyWE{
  C.~Holzhey, F.~Larsen and F.~Wilczek,
  ``Geometric and renormalized entropy in conformal field theory,''
Nucl.\ Phys.\ B {\bf 424}, 443 (1994).
[hep-th/9403108].
}

\lref\CalabreseRX{
  P.~Calabrese and J.~L.~Cardy,
  ``Time-dependence of correlation functions following a quantum quench,''
Phys.\ Rev.\ Lett.\  {\bf 96}, 136801 (2006).
[cond-mat/0601225].
}

\lref\CallanPY{
  C.~G.~Callan, Jr. and F.~Wilczek,
  ``On geometric entropy,''
Phys.\ Lett.\ B {\bf 333}, 55 (1994).
[hep-th/9401072].
}

\lref\BombelliRW{
  L.~Bombelli, R.~K.~Koul, J.~Lee and R.~D.~Sorkin,
  ``A Quantum Source of Entropy for Black Holes,''
Phys.\ Rev.\ D {\bf 34}, 373 (1986)..
}

\lref\PolchinskiTA{
  J.~Polchinski,
  ``String theory and black hole complementarity,''
In *Los Angeles 1995, Future perspectives in string theory* 417-426.
[hep-th/9507094].
}

\lref\MaldacenaDS{
  J.~M.~Maldacena and L.~Susskind,
  ``D-branes and fat black holes,''
Nucl.\ Phys.\ B {\bf 475}, 679 (1996).
[hep-th/9604042].
}

\lref\NozakiWIA{
  M.~Nozaki, T.~Numasawa and T.~Takayanagi,
  ``Holographic Local Quenches and Entanglement Density,''
[arXiv:1302.5703 [hep-th]].
}

\lref\AlmheiriRT{
  A.~Almheiri, D.~Marolf, J.~Polchinski and J.~Sully,
  ``Black Holes: Complementarity or Firewalls?,''
JHEP {\bf 1302}, 062 (2013).
[arXiv:1207.3123 [hep-th]].
}

\lref\HarlowTF{
  D.~Harlow and P.~Hayden,
  ``Quantum Computation vs. Firewalls,''
[arXiv:1301.4504 [hep-th]].
}

\lref\ParikhKG{
  M.~Parikh, P.~Samantray and ,
[arXiv:1211.7370 [hep-th]].
}

\lref\AzeyanagiBJ{
  T.~Azeyanagi, T.~Nishioka, T.~Takayanagi and ,
Phys.\ Rev.\ D {\bf 77}, 064005 (2008).
[arXiv:0710.2956 [hep-th]].
}



\Title{
\vbox{\baselineskip12pt
}}
{\vbox{\centerline{ Time Evolution of Entanglement Entropy  }\vskip .5cm
 \centerline{  from Black Hole Interiors }
}}

\bigskip
\centerline{Thomas Hartman and Juan Maldacena }
\bigskip

\centerline{ \it  School of Natural Sciences, Institute for
Advanced Study} \centerline{\it Princeton, NJ, USA}

\vskip .3in \noindent

We compute the time-dependent entanglement entropy of a CFT which starts in  relatively simple initial states. The initial states are the thermofield double for thermal states, dual to eternal black holes, and a particular pure state, dual to a black hole formed by gravitational collapse. The entanglement entropy grows linearly in time. This linear growth is directly related to the growth of the black hole interior measured along ``nice'' spatial slices. These nice slices probe the spacelike direction in the interior, at a fixed special value of the interior time. In the case of a two-dimensional CFT, we match the bulk and boundary computations of the entanglement entropy. We briefly discuss  the long time behavior of various correlators, computed via classical geodesics or surfaces, and point out that their exponential decay comes about  for similar reasons. We also present the time evolution of the wavefunction in the tensor network description.


 \Date{ }




\newsec{ Introduction }

A black hole interior is, almost by definition, difficult to probe.  Perturbatively,
an outside observer can never receive signals from behind the event horizon.
In a unitary quantum theory, however, information should not be lost.
According to the gauge gravity duality, all the relevant information is contained in the field theory that lives at the
boundary.

In this paper we discuss some observables whose gravity computation involves the interior.
 By `interior', we mean the region to the future of the event
 horizon heading towards the singularity, not to be confused with the second exterior region of an eternal black hole.  We consider both the eternal black hole, dual to a thermal state in CFT, and a particular
 black hole microstate that can be formed by gravitational collapse and has only one asymptotic region.

Our main tool is the entanglement entropy, computed holographically as the area of an extremal surface in $AdS$ ending on the boundary \refs{\RyuBV,\HubenyXT}.  In a static situation, extremal surfaces do not penetrate the event horizon \HubenyRY, but if the system is time dependent then it is possible to probe the interior.  A similar setup has been explored using the Vaidya spacetime, which describes a shell of null dust falling into a black hole   \refs{\HubenyXT,\AbajoArrastiaYT,\AparicioZY,\AlbashMV,\BalasubramanianCE,\BalasubramanianUR}.
However here we will use only the empty black brane geometry.

To understand how we introduce time dependence to probe the black hole interior, first consider an eternal black hole.  It is static under time evolution that runs forward on one side of the Penrose diagram, and backwards on the other. However, instead we will choose to time-evolve forward in both exterior regions, so the system is time dependent. This is a simple model for thermalization in the strongly coupled dual CFT.  The two-sided setup sounds artificial but in fact it is relevant to a more realistic thermalization process:  There is a class of black hole microstates that are similar  to the eternal black hole outside the horizon, but they lack the second asymptotic region.  These black holes are dual to time-dependent pure states in CFT undergoing thermalization.

Let us summarize the calculation.  The entanglement entropy of a quantum system is defined by separating the system into two parts, $A$ and $B$, on some fixed-time slice.  The reduced density matrix of subsystem $A$ is computed by tracing out other degrees of freedom, $\rho_A = Tr_B\ \rho_{total}$, and the entanglement entropy is $S_A = -Tr\rho_A \log \rho_A$. For $A$ a spatial region in a two-dimensional CFT, the entanglement entropy in the groundstate and a thermal state was computed in \refs{\HolzheyWE,\CalabreseEU} using the replica method.

Similar techniques were used in \CalabreseIN\ to compute the time-dependent entanglement entropy of a 2d CFT in a gapped excited state.  This is interpreted as modeling the
 time evolution after a global quantum quench, where we prepare the system in the groundstate of a gapped Hamiltonian $H_0$ and suddenly change $H_0 \to H$, so the system is no longer in the groundstate.  The system starts with only short-range entanglement. The result of \CalabreseIN\ is that the entanglement grows linearly, and eventually saturates at the thermal value after a time $t \sim L/2$ where $L$ is the size of region $A$.  That is, the subsystem eventually acts thermal due to correlations with the rest of the system. The state that is actually used in \CalabreseIN\ is given by starting from a boundary
 state in the CFT and evolving it in Euclidean time in order to render it normalizable. We call the resulting state ``the B-state''.  We will point out that the gravity
 dual of this procedure is very simple. It amounts to cutting the usual eternal black hole Penrose diagram in half by adding
  an end of the world brane in the bulk.

On the gravity side, as mentioned above,  the entanglement entropy is related to an extremal surface in $AdS$ that ends on the boundary of the region $A$ at the boundary. The proposal, made for static spacetimes in \RyuBV\ and extended to time-dependent backgrounds in \HubenyXT, is $S_A = {\rm Area}(\gamma)/( 4 G_N)$ where $\gamma$ is the extremal surface and $G_N$ is Newton's constant.  If there are multiple extremal surfaces ending on region $A$, the rule is to pick the one with minimal area. 

We take region $A$ to be half of space, or a finite strip of width $L$.  For the eternal black hole, $A$ consists of two identical pieces, one on each side of the black hole. For $L\gg \beta$ with $\beta$ the inverse Hawking temperature, the minimal area extremal surface at early times extends across the black hole from one asymptotic region to the other. As we evolve in time it progresses into the interior along special nice slices.  After a time $t\sim \beta$, the extremal surface gets `stuck' at a particular time in the interior, but continues to stretch along the spacelike direction of the interior.  This growth of nice slices along the spacelike direction in the interior is directly responsible for linear growth in entanglement entropy, $S_A = s v t$, where $s$ is the thermal entropy density and $v\leq 1$ measures the speed of entanglement growth.

If $A$ is a finite region of size $L$, then after $t \sim L/2$, the minimal area extremal surface jumps to a static surface sitting outside the horizon that reproduces the thermal entropy.  All of these statements also apply to the single-sided black hole microstate.

In two boundary dimensions, where the CFT calculation can be done explicitly, the gravity calculation gives linear growth with $v=1$.  For the eternal black hole we find an exact match to a thermal double CFT. In fact, when the region $A$ is half of
space, the result is fixed by conformal symmetry.
For the single-sided black hole our result agrees with the Cardy-Calabrese  B-state  \CalabreseIN.

The main conclusion is that the growth of the nice slices along the  spacelike $t$-direction in the interior is responsible for linear growth in entanglement entropy during thermalization.  We also discuss the relevance of this interior region for other observables, including correlators of heavy particles, strings, and branes.

In our setup the entanglement at $t=0$ is entirely short range\foot{ This differs slightly
 from the Vaidya calculations mentioned above,
 where the initial entanglement after the quench is spread over different scales.
 Other types of quenches have also been studied holographically in, for example, \refs{\TakayanagiWP,\AsplundCQ,\BasuFT,\BasuGG,\BuchelLLA,\BalasubramanianAT,\AllaisYS, \NozakiWIA}. We can view field theory in de-Sitter
 space also as a particular time dependent state. The growth of the entanglement entropy in that case \MaldacenaXP\  has features similar
 to those discussed here. }.
As the system evolves, entanglement spreads out over larger distances. This suggests a picture for the wavefunction of the CFT during thermalization, where the ordinary renormalization group
flow is supplemented by additional structure at scales larger than $\beta$.

The paper is organized as follows.  In section 2, we calculate the entanglement entropy from gravity in $AdS_{d+1}$. In section 3, we specialize to $AdS_3/CFT_2$, and compute the entanglement entropy from both bulk and boundary.  In section 4 we consider the region in the interior relevant to computing the long time behavior of
  the expectation values of local operators,   Wilson loops and higher dimension surface operators.  We comment on implications for the CFT wavefunction in section 5, and conclude with a discussion of the interpretation of the interior in section 6.

 \ifig\Penrose{(a) Penrose diagram of the maximally extended black branes we consider. We suppress the spatial
coordinates along the brane. There are two exterior regions $E_1$ and $E_2$, each of which has a boundary. The two copies
   of the corresponding field theories live at these boundaries. There are two interior regions $In$ to the future and $In'$ to the past. This can obtained from a Euclidean solution by cutting and pasting at a moment of time reflection symmetry, as
   indicated in (b). This procedure also produces the entangled state in the two copies of the field theory. It just corresponds to doing Euclidean time evolution over a time $\beta/2$.
   } {\epsfxsize4.5in\epsfbox{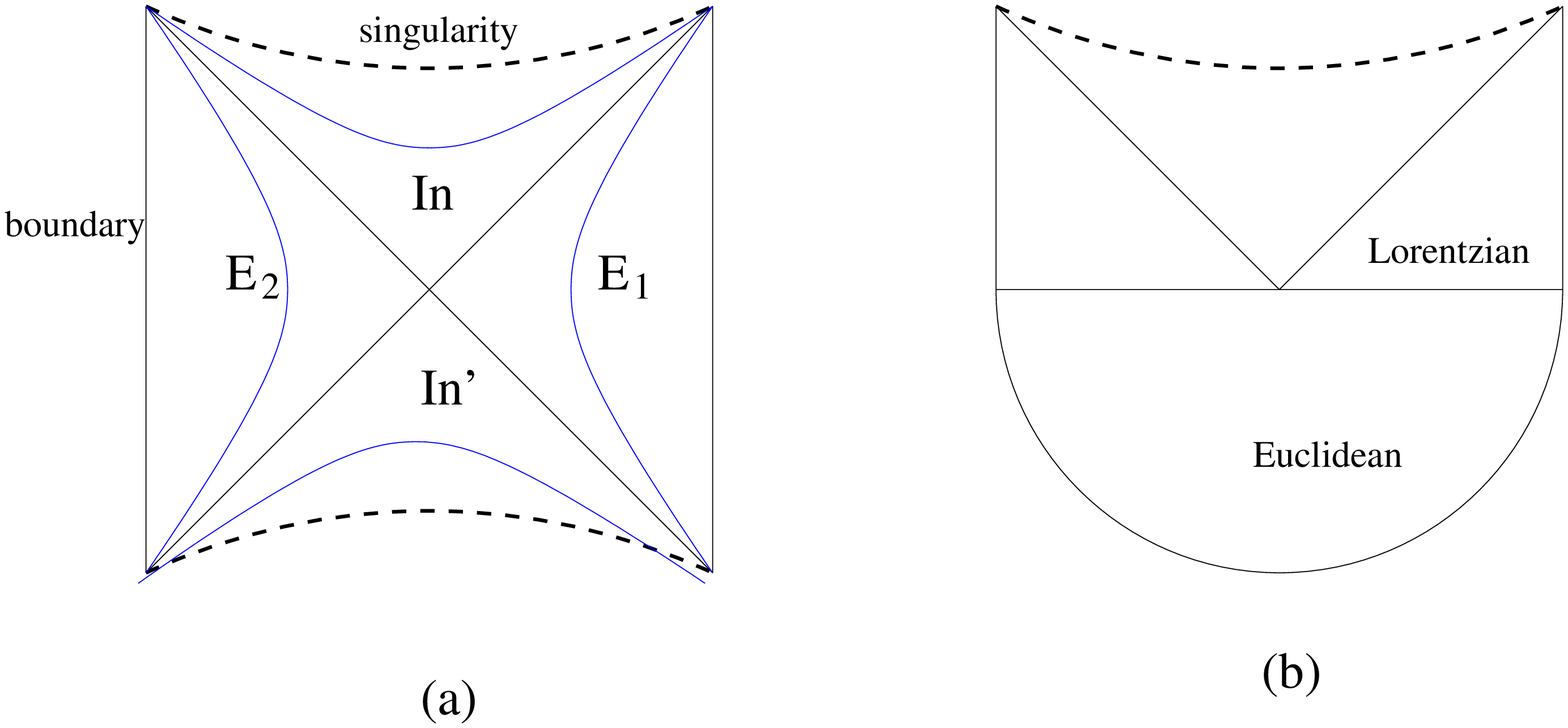}}

\newsec{ The growth of the entanglement entropy}\seclab\secbulk
We first consider black holes that correspond to the holographic duals of thermal field theories.
  We assume that we have one or more infinitely extended spatial directions, and consider a metric of the form
\eqn\metrf{
 ds^2 =  - g^2(\rho  ) dt^2 + h^2 (\rho ) dx_{d-1}^2  + d \rho^2 \ .
 }
We assume there is a horizon at $\rho =0$ and that for small $\rho$,
 \eqn\assumgh{
 g(\rho) =  \rho + o(\rho^3)~,~~~~~~~~~~h(\rho) = C + o(\rho^2) ~,~~(C\not =0)~~~ \ .
  }
 We set the temperature to  $\beta = 1/T = 2 \pi $, unless we explicitly  introduce $\beta$.
 We also assume that $g,h$ grow sufficiently fast for large
 $\rho $ so that we have a timelike boundary. Simple examples of such metrics are planar black
 branes in $AdS_{d+1}$ where (see appendix A)
 \eqn\gendi{
 h = { 2 \over d }
   \left( \cosh { d \, \rho \over 2 } \right)^{ 2/d} ~,~~~~~~~~~~ g = h \tanh  { d \, \rho \over 2 } \ .
 }
The full extended Penrose diagram is shown in \Penrose . (The particular case of
$AdS_3$ is special and will be treated in more detail in the next section).

The interior region corresponds to $ \rho = i \kappa $ and $t = t_{I} - i \pi/2$, so that $\rho e^{  t} $
is finite as we cross the horizon, and $t_{I}$ is real.
The coordinate $t_I$ is spacelike in the interior. Note that then $ - i g(i \kappa) $ is real and positive
in the interior.

The extended Penrose diagram also contains a second copy of the exterior.
The full spacetime is dual
to the thermofield double \refs{\IsraelUR,\MaldacenaKR}.
It has two boundaries which correspond to two copies of the field theory. These two copies are in an
entangled state of the form\foot{ The index $n$ is really continuous if we consider field theories in a
non-compact space. }
\eqn\entdouble{
 | \Psi \rangle = \sum_{n} |E_n \rangle_1 |E_n\rangle_2 e^{ - { \beta \over 2 } E_n } \ .
 }

We will now consider the entanglement entropy of a region ``$A$''.
Region $A$  consists of half of the space
of each of the two copies of the thermofield double. We will be separating each copy in two halfs at some
time $t_b$ which is the same on both sides. In other words, we take time to run forwards on both copies of
the field theory \foot{ Of course, running time forwards in one copy and backwards in the other is a symmetry of
the problem. Our time evolution is {\it not} a symmetry of the problem. }.
Running time forwards on both copies we are introducing some time dependence into the problem
and the entanglement entropy will depend on time. This corresponds to replacing $ e^{ - { \beta \over 2 } E_n }
\to e^{ - { \beta \over 2 } E_n - 2 i E_n t_b }$ in \entdouble .

The conjectured holographic recipe   for computing this entropy is  the following \refs{\RyuBV,\HubenyXT}.
We have to find the area of
an extremal codimension two surface  that ends at
the boundaries of the region\foot{This formula is an unproven conjecture, but it has been shown to be valid in
some special cases and has passed several checks \NishiokaUN .}.

\ifig\Geodesic{ (a) Extremal surface that computes the entanglement entropy. For large $t_b$, it sits very close
to a special critical spacelike surface in the interior. This region gives rise to the linear behavior in $t_b$ in
the entanglement entropy. (b) The same for the B-state. Here we can have the surface ending at the end of the
world brane.
   } {\epsfxsize5in\epsfbox{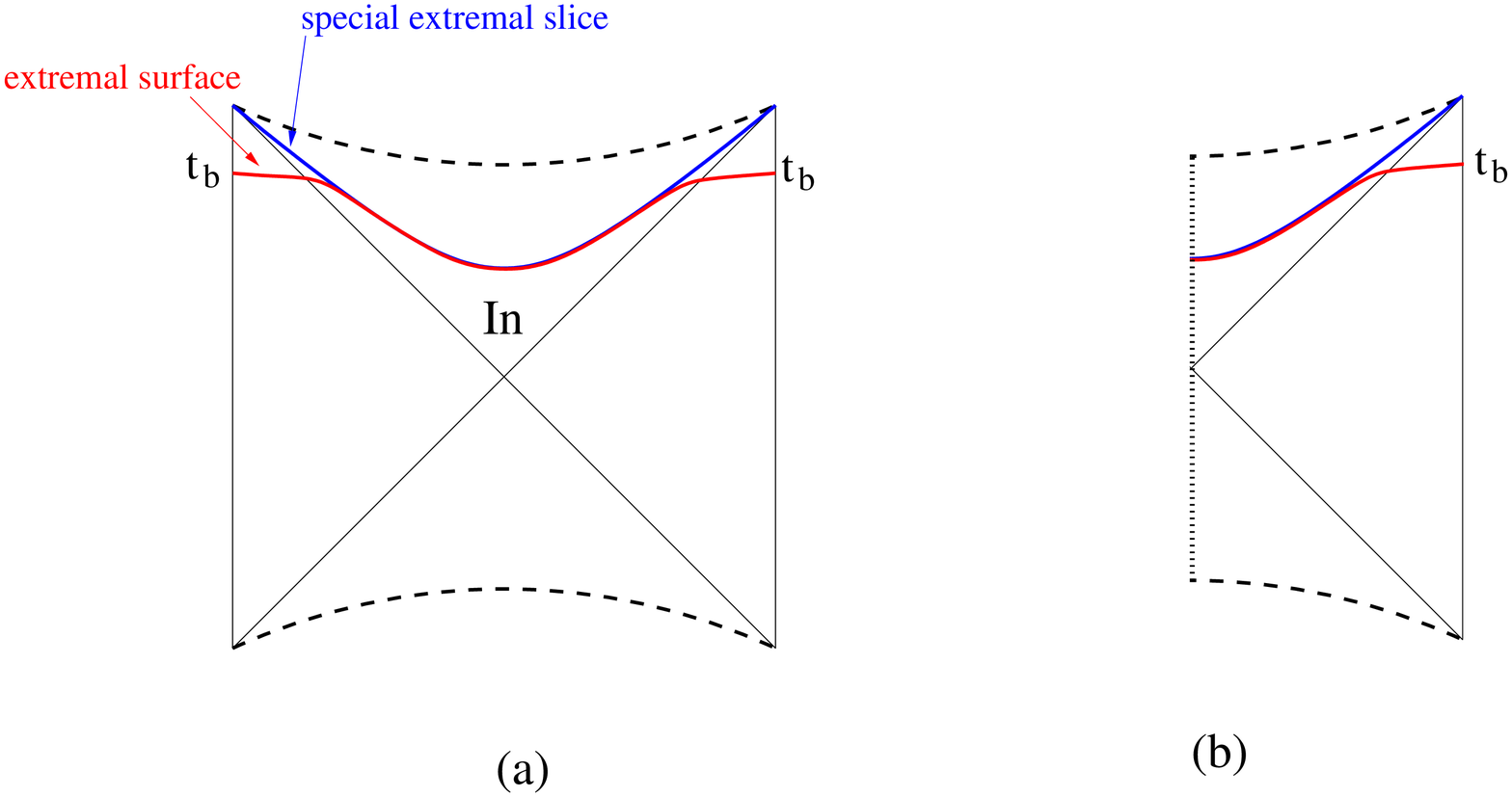}}

In our case, the symmetries of the problem suggest that if  region $A$ is
 the half space $x_1>0$, then the bulk
surface will be extended along $x_2, \cdots , x_{d-1}$ and will sit at $x_1=0$.
Thus finding the surface reduces to finding a function $t(\rho)$  (or $\rho(t)$) which describes how the surface
moves in the $t,\rho$ direction. In order to find it, we need to extremize the area
 \eqn\extra{
 A = V_{d-2} \int d t  \left[ h(\rho) \right]^{ d -2   } \sqrt{ - g^2(\rho) + \dot \rho^2 } \ .
 }
We first look for an extremal surface of the the form shown in \Geodesic . It has a $t_I \to -t_I$ symmetry
under the reflection that exchanges the two sides. So, at $t_I=0$ we expect that $\dot \rho =0$.

 It is easy to find the first integral  of the equation of motion of \extra .
 Since \extra\  is time independent,   the ``energy'' is
 conserved, giving
 \eqn\conseq{
 { g^2 h^{ d-2} \over \sqrt{ -g^2 + \dot \rho^2 }  } =    { - i g_0 h_0^{d-2}     }
 }
 where $h_0$ and $g_0$ are $g$ and $h$ evaluated at the interior point $\rho = i \kappa_0$ where $\dot \rho =0$. (Note that $g_0$ is purely imaginary).
  From here we can find
 \eqn\valrho{
  t(\rho) = - i { \pi \over 2 } -  \int_{i \kappa_0}^\rho  { d \rho' \over g \sqrt{ 1 - { g^2 h^{ 2 d -4}  \over g_0^2 h_0^{2d-4} }   } } \ .
}
\ifig\Contour{  Contour in the $\rho$ plane that gives $t(\rho)$. We avoid the pole at $\rho=0$ as indicated.
   } {\epsfxsize2.5in\epsfbox{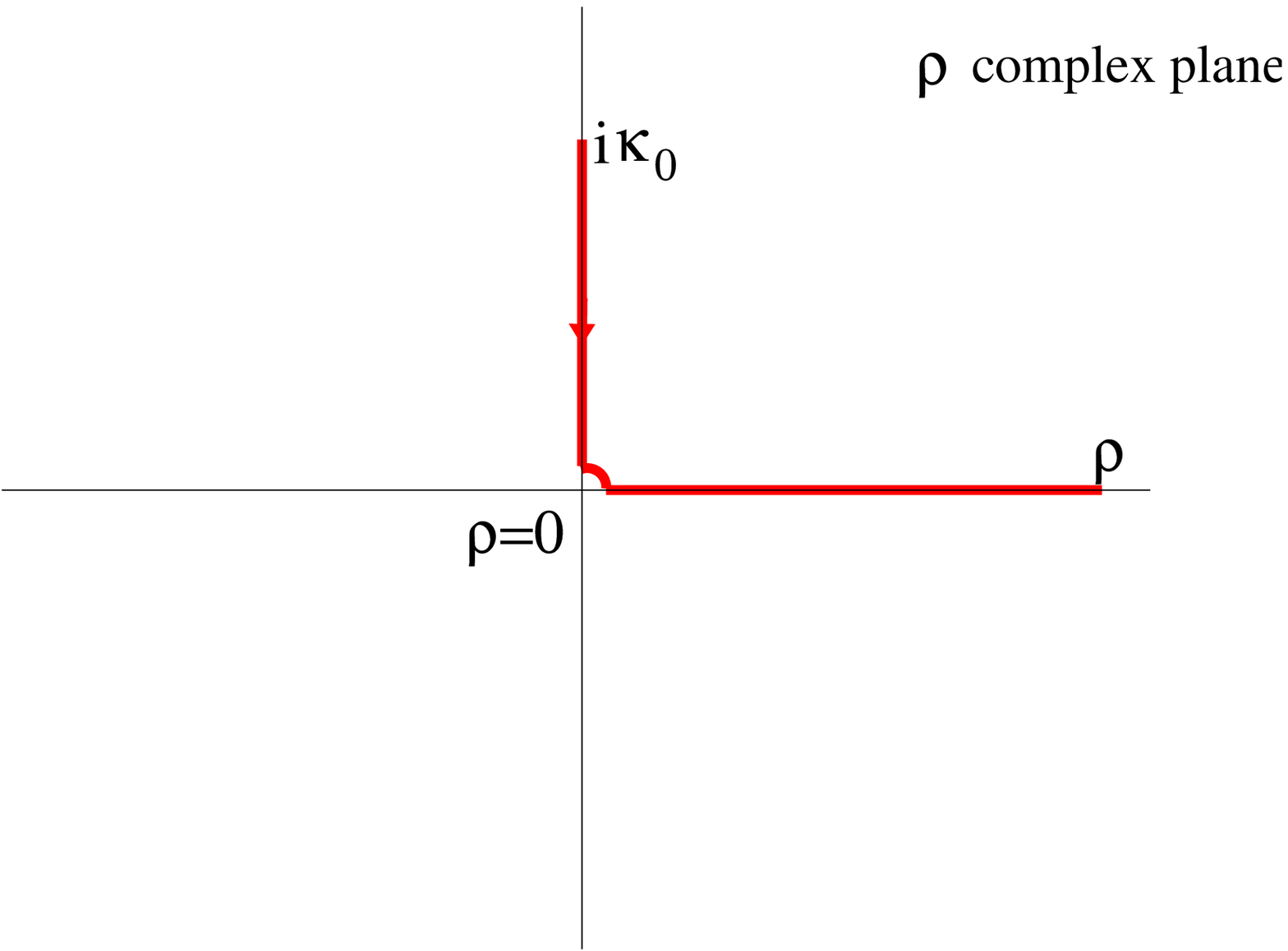}}

Note that the integral \valrho\
has a pole at $\rho =0$ and should be taken along the contour indicated in \Contour .
This pole at $\rho=0$  gives the expected behavior $t \sim - \log \rho $ as we cross the horizon, implying that
 $\rho e^t$ remains finite as the trajectory crosses the horizon.
 The value of the area is also simply given as
 \eqn\valare{
  A =  V_{d-2} \int dt h^{ d-2} \sqrt{ -g^2 + \dot \rho^2 }  =  2 V_{d-2}
  \int_{ i \kappa_0}^\infty d\rho   {    h^{  d -2} \over  \sqrt{1 - { g_0^2 h_0^{ 2 d - 4 }  \over  g^2 h^{ 2 d -4}} } } \ .
  }
 The factor of two arises from the symmetry of the configuration, see \Geodesic (a).

 In general this integral is finite as long as $\rho_0 = i \kappa_0$ is not exactly the point where
 $a \equiv ( -i g) h^{ d -2} $ is extremized.
 Let us denote the point where $a$ is extremized by $\rho_m = i \kappa_m$. Note that $a $ is zero at the horizon $
 \rho =0$ and then it starts growing as we go into the interior. It typically gets to a maximum value at $\kappa_m$ and then decreases again. This can be checked explicitly for \gendi .

 For finite values of $t_b = t(\rho=\infty)$ we find that $ \kappa_0 < \kappa_m$. But as $t_b$ becomes large, we find
 $\kappa_0 \to \kappa_m$ and that is the reason why $t_b$ in \valrho\ becomes large.
 For large $t_b$ the minimal area extremal surface has two regions. It lies at $\kappa \sim \kappa_m$ for a long
 range of the spacelike interior $t$-direction,  and then it crosses the horizon and goes to the boundary. The region of
 the solution where it crosses the horizon and goes near the boundary has a shape that depends on the
 details of the geometry but, for large $t_b$, it gives just a constant to \valare .
 On the other hand, the piece lying in the interior gives a large contribution linear in $t_b$ to the area.
 This linear contribution is computed by approximating the area by setting
 $\dot \rho =0$ and $g \to g_m$, $h \to h_m$ in the first equality in \valare . The form of the entanglement
 entropy for large $t_b$ is
 \eqn\larcof{
 S = { A \over 4 G_N}  = {  V_{d-2} \over 2 G_N}  \left[ a_m t_b + {\rm constant } \right] \ .
 }
 Of course $A$ in \valare\ is  divergent at large $\rho$, but the divergence is independent of $t_b$. The coefficient $a_m = \sqrt{ -g^2_m} h_m^{d-2} $ is the value  of $a$
 at the extremum, which is a maximum. 
 These surfaces were also considered in \AzeyanagiBJ , here we are expanding on their
 interpretation. 

 It is interesting to express this number in units of the entropy density,
 $ s =  { h(0)^{d-1}  \over 4 G_N} $.  This
 gives us a measure of the speed of growth of the entropy as
 \eqn\speedgr{
   { \partial S \over \partial_{t_b} }  = v V_{d-2} s ~,~~~~~~~~
   v =  { \sqrt{-g^2_m} h_m^{d-2} \over h(0)^{ d -1} } \ .
  }
  For the $AdS_{d+1}$ black brane  metrics \gendi \ we find that
  \eqn\figco{
  \tan { d \kappa_m \over 2 }
   = \sqrt{ d \over d -2 } , ~~~~~~~ v = { \sqrt{ d }  (d-2)^{ {1\over 2 } -{ 1\over d }  }  \over  [ 2( d -1)]^{ 1 - { 1 \over d } }  } \ .
  }
 For $d=2$ $v=1$, for $d=3$ $v$ is slower than  the speed of sound  and for $d>3$ $v$ is
  larger than the speed of sound and always bigger than $1/2$.

\subsec{ Other shapes for region $A$ }

Let us now discuss what happens if we take the $A$ region to have different shapes.
A simple shape is  $I \times R^{d-2}$ on each of the two copies of the field theory, where $I$
is an interval of length $L \gg \beta $. The minimal area surface consists of two disconnected surfaces, each
of the form discussed above. Thus we get the same answer, up to a factor of two.
This is the correct answer for   times smaller
than $L$. When the time becomes of order $L$ then there is another extremal surface whose area becomes smaller.
It consists of two disconnected surfaces, each staying outside the horizon, but close to the horizon.
For large $L$ their contribution goes like the black hole entropy density times the length of the interval.

More generally, consider a region $A$ given by the union
of two identical regions $R$ in each of the copies of the field theory.
We take $R$ to be   very big compared to $\beta$ (and assume that its boundary is
 not varying rapidly on distances
smaller than its size).
 Then we get the same result as in \larcof\ with $V_{d-2} \to B_{d-2}$ where $B_{d-2}$
is the {\it area}  of the boundary of region $R$.

Notice that the thermofield double \entdouble\ (at $t=0$)
is behaving as a state with a mass gap from the point of
view of the two copies of the field theory. Here what we mean is that all equal time 
correlation functions decay exponentially as we separate them in space. Also the entanglement entropy for the 
regions in the above paragraph goes like the area of their boundaries. 
 Of course the system is not gapped as a lorentzian solution, 
where we can have very low energy excitations by considering particles very near the horizon. 

When $t_b$ becomes comparable to the size of the region $R$,  we can have other surfaces which can give a
smaller area. These are surfaces that stay on each of the exterior regions. The surfaces are disconnected, they
live purely at $t_b$ and do not cross the horizon. However, they asymptote to the horizon and, for large regions,
they give an entropy of the form $S = 2 V_{R} s $, where $s$ is the entropy density of the black brane
and $V_R$ is the {\it volume} of the  region $R$.

\subsec{ Black holes formed from pure states }\subseclab\seconesidedbh

\ifig\Penrose{(a) Penrose diagram for the B-state. We have an end of the world brane or orbifold plane
 along the dotted line. This breaks the time translation symmetry. The corresponding state can be produced by cutting a
 Euclidean solution at a moment of time reflection symmetry, as indicated in (b). Note that in the Euclidean solution the
 end of the world brane reaches the boundary. Thus the Euclidean field theory has a boundary. The state at $t=0$ corresponds
 to performing Euclidean time evolution on the conformally invariant boundary state by an amount $\beta/4$ of Euclidean time.
   } {\epsfxsize4.5in\epsfbox{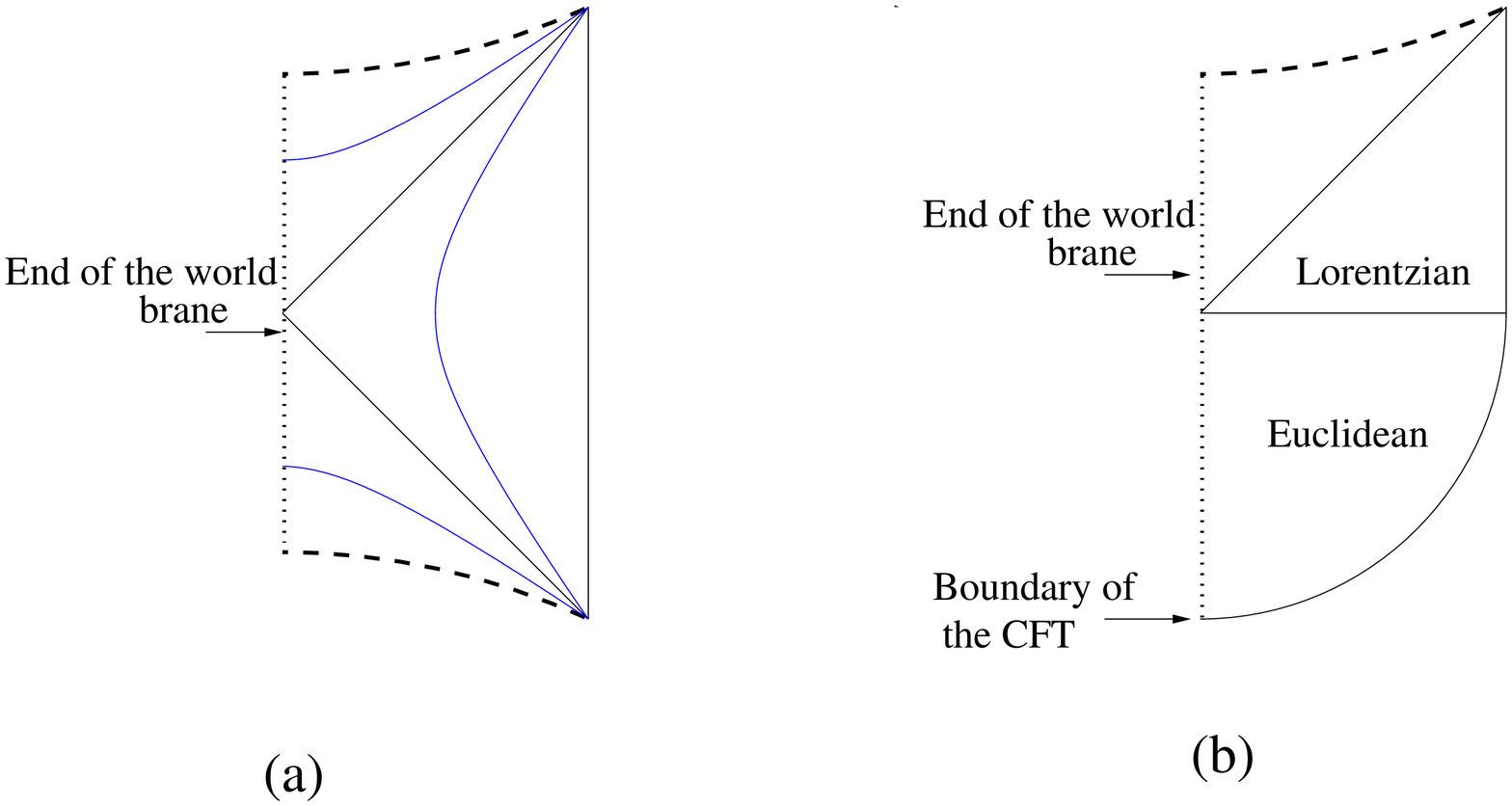}}

The behavior of the entanglement entropy under black hole formation was studied in numerous publications which
found features which are the same as what we discuss here
\refs{\HubenyXT,\AbajoArrastiaYT,\AparicioZY,\AlbashMV,\BalasubramanianCE,\BalasubramanianUR,%
\AsplundCQ,\BasuFT,\BasuGG,\BuchelLLA,\BalasubramanianAT,\AllaisYS}. Several of these publications focused on the case
of Vaidya   metrics. Here we would like to discuss a very simple black hole formation situation which is
  related to the above  discussion. In addition, it is the precise holographic dual of the idealized
quench configurations discussed in \CalabreseIN.

The solutions are simply  the eternal black hole solutions with an end of the world brane that cuts them in
half. This end of the world brane can have different physical origins. It could be a true end of the world brane
in an M-theory solution, or it could arise from various types of orbifolds. Some examples are given in \MaldacenaKR \foot{
An example of an end of world brane is the end of the world brane of M-theory. We can get others from type I' theory or
various orbifolds. We do not want to consider here the ``geon'' configuration discussed in \LoukoHC , because
those solutions involve a boundary condition which is not local in the spatial coordinates of the CFT (even though they
are perfectly well defined and reasonable objects to study).}.

This is a state whose description is almost as simple as the one above. Namely, we consider a field theory
with a boundary condition. We then evolve in euclidean time for a time $\beta/4$,
\eqn\solev{
 |\Psi \rangle = e^{ - { \beta \over 4 } H } | B \rangle \ ,
 }
 where $|B\rangle$ is a boundary state in the conformal field theory.
 We can start with this state at $t=0$ and evolve it in Lorentzian signature.

The state at $t=0$ has a mass gap of the order of the temperature\foot{
Again, this is a mass gap in the sense that the 
equal time correlation functions decay exponentially in space. Also, the entanglement entropy of a region goes like
the area of the region. Of course, the lorentzian spacetime can support very low energy excitations which are obtained by
placing particles very near the horizon.
 }.
This is a common feature of (stable) AdS black holes.

The holographic computation of the entanglement entropy is essentially identical to the one we had above.
The extremal surface used to compute the entropy can end on the end of the world brane. This end of
the world brane sits at $t_I =0$. So the solutions are the same as the ones we considered above.
And the entanglement entropy is essentially the same, up to a factor of two, since we now consider only
one of the sides.

We see that as time evolves the entanglement entropy grows.
This growth of the entanglement entropy can be understood from two points of view.
From the field theory point of view, the local entanglement that existed at $t=0$ is spread out by the
time evolution.
Thus, for large times the entanglement is highly non-local along the non-compact directions
of the field theory and it is spread over a region of size $t_b$.

In the bulk solution the entanglement is given by a surface that starts from the boundary and ends on the
end of the world brane. See \Geodesic (b). The linear growth in $t_b$ is due to the fact that the extremal surface
is growing along the $t_I$ direction inside the horizon.
The position along the timelike direction in the interior
is barely changing.

The shape of the extremal surfaces is essentially identical to the shape of the so called
``nice slices'' that are usually introduced in discussions of black hole
information loss \PolchinskiTA .
The growth of the entanglement entropy is due to the growth of these nice slices. These slices grow because
we keep adding space in the interior, making the range of $t_I$ bigger.

  \ifig\figcyl{  Euclidean cylinder with twist operator insertions used to compute the entanglement entropy $S_A$ in a finite-temperature CFT, where region $A$ is (a) the half-line, and (b) a finite interval.
   } {\epsfxsize 
   12cm
   \epsfbox{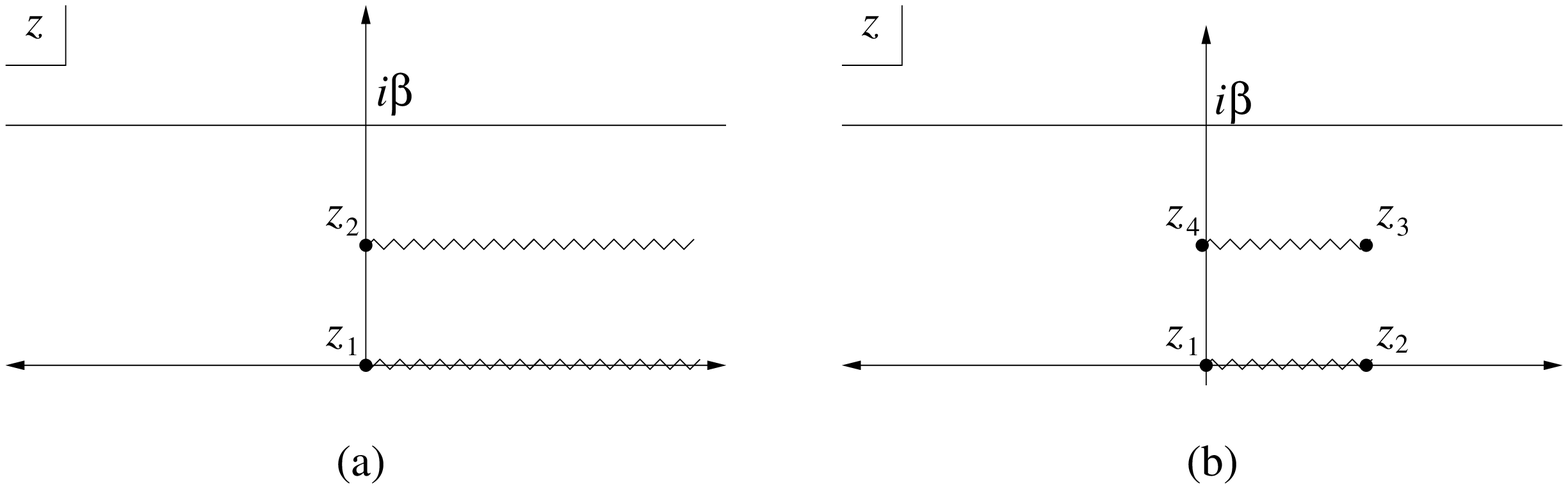}
   }

\newsec{ BTZ black strings and two dimensional CFT}

Here we consider a two dimensional CFT on an infinite spatial line.
 In this case the entanglement entropy can be computed in CFT using the replica technique
 \refs{\CallanPY,\HolzheyWE,\CalabreseIN,\CalabreseEU,\CalabreseRG,\CalabreseQY}  and can be  matched quantitatively to the gravity formula.
 We will also describe the gravity calculation in the previous section from an alternative viewpoint,
  using the fact that the `planar' black hole in $AdS_3$ is just ordinary $AdS_3$ written in different coordinates.

We consider four scenarios, both in gravity and in CFT: the subregion is either a half-line or a finite interval, and the CFT is either in a thermal state or in the pure B-state.

\subsec{CFT at finite temperature: half-line}\subseclab\cfttwoA
First we consider a finite-temperature CFT and choose region $A$ to consist of the half-line $x >0$, both in the CFT and in its thermal double. In Euclidean signature, insertions in the CFT are at Im $z=0$ and insertions in the thermal double are placed at Im $z=i\beta/2$. At time $t=0$, the $n^{th}$  power of the
reduced density matrix, $\rho_A^n$,
 is a Euclidean path integral on an $n$-sheeted cylinder with $z \sim z + i \beta$, shown in \figcyl(a).   This is given by a two-point function of twist fields inserted at the branch points. The answer at time $t$ is computed by analytically continuing the positions of the twist fields to Lorentzian time,
\eqn\cfthalfa{
\eqalign{
I_n &\equiv \tr\, \rho_A^n = \langle \Phi^+(z_1,\bz_1) \Phi^-(z_2,\bz_2) \rangle  \cr
z_1 &= \bz_1 = 0 \ , \quad\quad z_2 = -\bz_2 =  2t + i \beta/2 \ .
}}
Note $z_2^* \neq \bz_2$ because of the continuation to Lorentzian $t$. We have chosen time to run forward in both copies of the CFT (giving the factor of 2), so it is not a symmetry of the thermal state. The twist fields have dimension
\eqn\twistdim{
\Delta_{n} = \frac{c}{12}\left(n - \frac{1}{n}\right) \ ,
}
where $c$ is the central charge.

Mapping to the plane  $w = e^{2\pi z/\beta}$ gives the two-point function
\eqn\twopae{
I_n = \left[ {\beta\over 2\pi \epsilon}{\cosh\left(2\pi t\over \beta\right)}\right]^{-2\Delta_n} \ ,
}
where the short-distance cutoff $\epsilon$ comes from regulating the twist fields. Therefore the entanglement entropy $S_A = -\partial_n I_n|_{n=1}$ is
\eqn\cftha{
S_A = {c\over 3}\log\left(2\cosh {2\pi t\over \beta}\right) + 2 S_{\rm div} ~,~~~~~~~~~~~
 S_{\rm div} = {c\over 6 }\log\left(\beta\over 4\pi \epsilon\right) \ .
}
We see that for $t\gtrsim \beta$, the entanglement entropy grows linearly in time.

\subsec{CFT at finite temperature: finite interval}\subseclab\cfttwoB
Now take region $A$ to consist of two identical intervals of length $L$, one in the CFT and one in the thermal double. The case $t=0$ was discussed in \MorrisonIZ\ and is drawn in \figcyl(b). At finite time, the reduced density matrix is given by a 4-point function of twist fields,
\eqn\zerto{\eqalign{
I_n  = &\langle \Phi^+(z_1,\bz_1)\Phi^-(z_2,\bz_2)\Phi^+(z_3,\bz_3) \Phi^-(z_4,\bz_4) \rangle \ , \cr
&\quad z_1 = \bz_1 = 0 \ , \quad z_2 =\bz_2 = L , \cr
z_3 = 2t +L+ &i \beta/2 , \quad \bz_3 = -2t + L - i\beta/2 , \quad z_4 = -\bz_4 = 2t + i \beta/2 \ .
}}
Mapping to the plane, this becomes
\eqn\betteri{
I_n = \left( { \beta \over 2 \pi }  \right)^{-4\Delta_n} \left( 2 \sinh { \pi L \over \beta } \right)^{ - 4 \Delta_n }
\left( x_{\rm cross} \bar x_{\rm cross} \right)^{\Delta_n}
G_n(x_{\rm cross}, \bar{x}_{cross})
}
where $x_{\rm cross}$ is the conformal cross-ratio
\eqn\crossrat{
x_{\rm cross} = {(w_1 - w_2)(w_3 - w_4)\over (w_1 - w_3)(w_2 - w_4)} = {2\sinh^2(\pi L/\beta)\over \cosh(2\pi L/\beta) + \cosh(4\pi t/\beta)} \ ,
}
and
\eqn\planeg{
G_n(w, \bar{w}) = \langle \Phi^+(0)\Phi^-(w, \bar{w})\Phi^+(1)\Phi^-(\infty)\rangle_{plane} \ .
}
We also see that $x_{\rm cross} = \bar x_{\rm cross} $.
At this point we have mapped the problem to the 4-point function of twist correlators on the plane, which is the groundstate entanglement entropy of two disjoint segments in a single CFT.  The extra factors in \betteri\ come from the fact that the two intervals are boosted.  The problem on the plane in theories with a gravity dual was studied in detail by Headrick \HeadrickZT, building on previous work on multiple intervals in \refs{\HeadrickKM,\HubenyRE}. Therefore our calculations are just a small variation on \HeadrickZT\ with a different interpretation because now the cross-ratio depends on time.

At early times $\beta \ll t \ll L/2$, we see that we are doing the OPE around $x_{\rm cross} \sim 1$ and the correlator factorizes,
\eqn\ansearly{
\eqalign{
I_n &\approx \langle \Phi^+(z_1,\bz_1) \Phi^-(z_4,\bz_4) \rangle \langle \Phi^+(z_2,\bz_2)\Phi^-(z_3,\bz_3)\rangle \cr
 &=
  \left( { \beta \over 4 \pi \epsilon}  \right)^{-4\Delta_n} e^{-8\pi \Delta_n t/\beta} \ .
}}
Because of the exponential dependence on $ t,L$, this formula is actually valid until $t$ is within
 $\beta$  of $L/2$. We see that we get twice the answer for the half line in \cfthalfa , for any CFT.
  At late times $ t \gg L/2 $, the correlator factorizes in the other channel $x_{\rm cross} \sim 0$,
\eqn\anslate{
I_n \approx
 \left( { \beta \over 4 \pi \epsilon}  \right)^{-4\Delta_n}
e^{-4\pi \Delta_n L /\beta} \ .
}
From these expressions we find the entanglement entropy
\eqn\anscft{
S_{A} = \left\{
\eqalign{
& 4 S_{\rm div}  + {4\pi c t \over 3\beta} \quad\quad t \lesssim L/2
\cr
&4 S_{\rm div}  + {2\pi c L\over 3\beta}\quad\quad t \gtrsim L/2
}
\right. \  \  \ ,
}
with $S_{\rm div} $ in \cftha .

That is, the entanglement entropy grows linearly until it saturates at the thermal value. Generically, the transition between the two regimes is smoothed out over a time of order $\beta$. In theories with gravity duals we get a sharp transition, discussed below.

\subsec{CFT in the  pure  B-state }\subseclab\cftpure
Now consider a CFT in a pure state
\eqn\bstate{
|\Psi\rangle = e^{-\beta H/4}|B\rangle \ ,
}
where $|B\rangle $ is a boundary state.
Correlation functions in the state $|\Psi\rangle$ are computed by a Euclidean path integral on a strip,
\eqn\imz{
0\leq {\rm Im}\ z \leq {\beta \over 2} \ ,
}
with the boundary condition $B$ imposed at the top and bottom of the strip. Therefore the entanglement entropy of a half-line at time $t$ is related to the one-point function on the strip,
\eqn\twistone{
I_n = \langle \Phi^+(z_1, \bz_1) \rangle \ , \quad z_1 = -\bz_1 = t+i\beta/4 \ .
}
By mapping to the upper-half plane $w = e^{2\pi z/\beta}$ where the conformally invariant one-point function is $( w - \bar w )^{-\Delta_n}$, we find
\eqn\uhpone{
I_n = \left( {\beta\over 2 \pi \epsilon} \cosh \left(2\pi t \over \beta\right)\right)^{-\Delta_n} \ .
}
Therefore the entanglement entropy of the half line, $S_A = -\partial_n I_n|_{n=1}$, is
\eqn\quhe{
S_A = {c\over 6}\log\left(2  \cosh {2\pi t\over \beta}\right) + S_{\rm div}  \ .
}
This is half of the answer obtained in \cftha\ for the half-line in a thermal state. The reason for this is clear: in the pure B-state, when we compute the correlator on the upper-half plane, there are image points in the lower-half plane.  The image points play the same role that insertions in the thermal double played in the previous calculation.  The same is true for the finite interval (see \CalabreseIN\ for details of the computation) in the two OPE limits that we considered.  In summary, we have in both cases
\eqn\sumq{
S_A^{\rm{pure\ B-state }} = {1\over 2} S_A^{\rm{thermal\ double}} \ ,
}
for the linear rise and, for the interval, the late-time constant behavior.

   \ifig\figinnerbtz{  Left: Bulk and boundary regions of the BTZ black string.  The boundary regions I and III are the two Rindler wedges, and region II is the Milne patch.  The exterior of the black string ends on region I and extends perpendicularly into the bulk. The shaded region in the bulk is the interior, which reaches the boundary only along the light cone $x_0^2-x_1^2 = 0$. The future bulk region lies to the future of the shaded region and its boundary is the Milne patch.
   Right: Endpoints of the entanglement region.
   } {\epsfxsize
 12cm
    \epsfbox{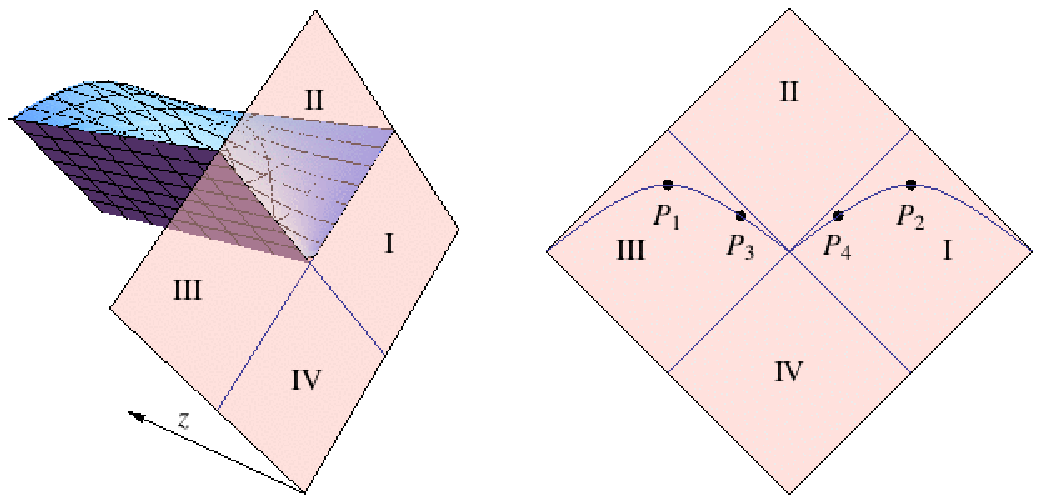}
   }

\subsec{BTZ black string}

The BTZ black hole is a quotient of $AdS_3$ \BanadosWN.  The results of section \secbulk\ apply to black holes with an infinite spatial direction, so let us unwrap the BTZ angle, giving back ordinary $AdS_3$.  Obviously now there is only one boundary. When we unwrap the quotient, the two boundaries of the BTZ Penrose diagram become the two wedges of Rindler space in the CFT \MaldacenaBW\ 
(see \ParikhKG\ for a more recent discussion).
 This maps the problem of computing extremal surfaces in section \secbulk\ to a very simple one, so we can find the extremal surfaces explicitly and see exactly how they pass through the bulk.

First we will describe the various regions of the black string.  It is useful to relate each coordinate patch to Poincar\'{e} coordinates,
\eqn\btzpoin{
ds^2 = \frac{1}{z^2}(-dx_0^2 + dx_1^2 + dz^2) \ ,
}
which ends on the Minkowski diamond of the CFT, pictured in \figinnerbtz. The exterior metric of the black string is
\eqn\btzc{
Exterior: \quad ds^2 = -\sinh^2\rho dt^2 + \cosh^2\rho dx^2 + d\rho^2 \ .
}
These coordinates cover a portion of the Poincar\'{e} patch. Near the boundary, they are related by
\eqn\btzchange{
x_1\pm x_0 \approx   e^{-x \pm t} \ , \quad\quad  {1\over z} \approx {1\over 2} e^{\rho - x} \ ,
}
so we see that $t,x$ cover one Rindler wedge of the Minkowski diamond. This is region I in  \figinnerbtz. We can reach the other Rindler wedge (region III) by continuing $t \to t + i \pi$, which is the same continuation that would take us from one side of the black brane Penrose diagram to the other.

The black string metric \btzc\ has a horizon at $\rho = 0$.  To go behind the horizon we continue $\rho = i\alpha, t = \tilde{t} - i\pi/2$, and find the interior metric
\eqn\btzin{
Interior: \quad ds^2 = \sin^2\alpha d\tilde t^2 + \cos^2\alpha dx^2 - d\alpha^2 \ .
}
These coordinates cover the shaded interior region in \figinnerbtz, which meets the boundary only along the light cone $x_0^2-x_1^2 = 0$.  Note that the interior of the black string does not correspond to the future diamond of the Milne
 parameterization of flat space, region II. That bulk region, to the future of the shaded portion in \figinnerbtz, can be reached by setting $\alpha = \pi/2 - i \tilde{\rho}, x = \tilde{x} - i \pi/2$, so
\eqn\btzfuture{
Future: \quad ds^2 = \cosh^2\tilde{\rho}d\tilde t^2 - \sinh^2\tilde\rho d\tilde x^2 + d\tilde\rho^2 \ .
}
The three bulk regions that we have just described --- exterior, interior, and future --- are related to Poincar\'{e} coordinates in appendix B.  The boundaries of the different regions are
\eqn\boundsregions{\eqalign{
Exterior: &\quad x_0^2 - x_1^2  \leq 0 \cr
Interior: &\quad 0 \leq x_0^2 - x_1^2 \leq z^2\cr
Future: &\quad  z^2 \leq x_0^2 - x_1^2  \ .
}}
There are actually two interior patches, inside the past and future horizons; the coordinates \btzin\ cover the patch with $x_0>0$.

\subsec{Holographic  entanglement: half-line}

Now, consider a subregion $A$ that consists of the half-space $x > 0$ in each of the two Rindler regions, at time $t_b$.  We will apply the holographic entanglement formula to compute $S_A$. In Poincar\'{e} coordinates $(x_0,x_1)$, the endpoints of region $A$ in the left and right Rindler wedges are
\eqn\plr{
P_{1} = (\sinh t_b, -\cosh t_b) \ , \quad P_2 = (\sinh t_b, \cosh t_b) \ .
}
The two points are at equal times and separated by $\Delta x_1 = 2 \cosh t_b$, see \figinnerbtz. Note that the computation of the entanglement in a CFT and its thermal double has been mapped to the problem of a single interval in a single CFT.  The geodesic that joins $P_1$ and $P_2$ is just a semicircle, given by the equations
\eqn\semic{
z^2 + x_1^2 = \cosh^2t_b  , \quad\quad x_0= \sinh t_b \ .
}
We will follow this semicircle through the various regions of the black string. In the coordinates \btzc, these equations imply $x = 0$ and
\eqn\semirho{
\sinh t \tanh\rho = \sinh t_b \ .
}
We see that as $\rho$ decreases, $t$ increases and goes to infinity at the horizon.  In the interior \btzin ,
 the semicircle obeys
\eqn\semiin{
\cosh\tilde t \tan\alpha = \sinh t_b \ .
}
As $\alpha$ increases from zero, $\tilde t$ decreases until $\tilde t = 0$ where
it meets the curve coming from the other Rindler wedge. If $t_b \gg 1$, then $\alpha$ will be very close to $\alpha_m = \pi/2$ for most of the range $\tilde t \lesssim t_b$. If we had performed the BTZ quotient, this special slice at $\alpha_m$ would be the  singularity of the BTZ  black hole.

We compute the length of the semicircle in Poincar\'{e} coordinates,
\eqn\aselij{
A_{12} = 2\cosh t_b \int_0^{\cosh t_b} {dz\over z}{1\over \sqrt{\cosh^2t_b - z^2}} \ .
}
The divergence near the boundary is regulated by placing a cutoff at $\rho = \rho_{max}$. Defining $\epsilon=e^{-\rho_{max}}$ and using \btzchange, the integral is cut off at $z=2\epsilon$. The final answer for the entanglement entropy is then
\eqn\singa{
S_A = \half \log \left(\cosh t_b \over \epsilon\right)\ .
}
Reintroducing the temperature $\beta=2\pi$ and Newton's constant $G_N = {3\over 2c}$, this becomes
\eqn\singatwo{
S_A = {c\over 3} \log \left(2 \cosh{ 2\pi t_b\over \beta}\right) +  2 S_{\rm div} \ ,
}
in agreement with the CFT calculation \cftha.

\subsec{Holographic entanglement: finite interval}

Now let us choose region $A$ to be an interval of length $L$ on each side of the Rindler diagram.  The endpoints are shown in \figinnerbtz. In Minkowski coordinates, we see that this is just the calculation of the entanglement entropy for  two boosted intervals.

\ifig\figconfigs{ Geodesics that connect the endpoints of the entanglement region, projected onto the $x_1,z$ plane.  There are two choices, shown as dashed or dotted lines.  Dashed geodesics go through the interior and dominate at early times.  Dotted geodesics stay within each exterior region and dominate at late times.
   } {\epsfxsize
8cm   \epsfbox{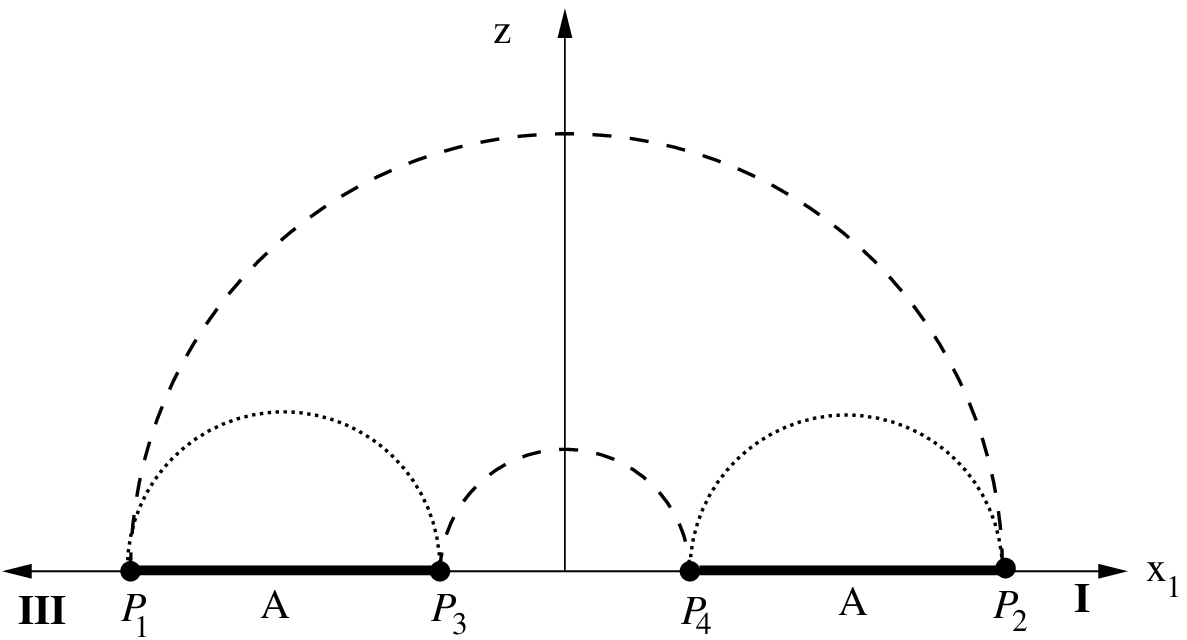}
   }

There are two ways to draw the bulk region ending on region $A$, as shown in \figconfigs.  (The figure is projected onto the $x_1,z$ plane; since the intervals are boosted, it is not a single time-slice.) Assuming $t_b \gg \beta$, the minimal length surface at early times consists of two geodesics that cross from one Rindler wedge to the other: one from $P_1$ to $P_2$ that was described above, and one joining the points
\eqn\pmore{
P_{3} = e^{-L} P_1  , \quad\quad P_4 = e^{-L} P_2 \ .
}
These are the dashed lines in \figconfigs.  The two geodesics have the same regulated length, because the $L$-dependence disappears when we use the appropriate cutoff $\rho = \rho_{max}$, $z = 2\epsilon e^L$. Therefore we just multiply \singa\ by two, and find
\eqn\repc{
S_A^{ (1)} = {4\pi c \over 3\beta}t_b +  4 S_{\rm div}
}
in agreement with the CFT result \anscft\ at early times.

The other configuration of geodesics ending on region $A$ (drawn as dotted lines in \figconfigs) is to connect $P_1$ to $P_3$ and $P_2$ to $P_4$. These geodesics do not pass through the interior. Instead they skim the outside of the horizon, so we expect an answer related to the horizon area, i.e., the thermal entropy. It is easiest to work in exterior coordinates \btzc. The symmetries of this configuration allow us to shift $x \rightarrow x - L/2$ so that the geodesics lie in equal-$t$ slices, then set $t = 0$ because the metric is static. Now we compute the distance between
\eqn\newpp{
P_2' = (e^{-L/2}, 0) ,\quad P_4' = (e^{L/2}, 0) \ .
}
The geodesic is again a semicircle in Poincar\'{e} coordinates and its length is
\eqn\slie{
A_{24} = \cosh\left(L\over 2\right)\int {dz\over z}{1\over\sqrt{ \cosh^2(L/2) - z^2}} \ ,
}
where the integral is over two segments: one from $z_{min} = 2\epsilon e^{L/2}$ to $z_{max} = \cosh(L/2)$, and one from $z_{min} = 2\epsilon e^{-L/2}$ to the same $z_{max}$. Reintroducing $\beta, G_N$ as above and assuming $L \gg \beta$, the entanglement entropy $S_A^{(2)} = (A_{13} + A_{24})/4$ is
\eqn\entlate{
S_{A}^{(2)} = {2\pi c\over 3\beta}L + 4 S_{\rm div} \ .
}
Comparing to the other configuration \repc, we see that indeed the minimal area surface is the connected diagram at early times and the disconnected diagram at late times.  Putting everything together, we have linear growth in the entanglement entropy given by $S_A^{(1)}$ until $t_b = L/2$, followed by a sharp transition to the thermal value $S_A^{(2)}$.  This agrees with the CFT result \anscft , but now the transition is sharp.

The fact that there is a sharp transition rather than a smooth cross-over in CFTs with a gravity dual was understood in  \HeadrickZT. The configuration in \HeadrickZT\ is two disjoint intervals $[0,x_{\rm cross}]\cup [1,\infty]$ at fixed time, and the sharp transition appears at $x_{\rm cross}=\half$. It is argued that this transition is related to the Hawking-Page phase transition between thermal AdS and the BTZ black hole (and higher-genus analogues of this transition). To see this, note that in the case $n=2$, the replica CFT lives on a torus, so sharp transitions in the torus partition function can be expected to lead to sharp transitions in the entanglement entropy. (See \HeadrickZT\ for details.)  In our case the intervals are boosted, and the cross-ratio in \crossrat\ is a function of time. The transition point $x_{\rm cross} = \half$ corresponds to $t = L/2$, so this becomes the transition between $S_A \sim  t$ and $S_A \sim $ constant.

\subsec{Holographic entropy of the pure B-state }

In the case of the single sided black string described in section \seconesidedbh, the answer for the entanglement entropy \singatwo, \repc, or \entlate\ is simply divided by two.  The holographic dual of the single sided black string is time evolution in a pure state $|\Psi\rangle = e^{-\beta H/4}|B\rangle$, which is precisely the Cardy-Calabrese calculation described in section \cftpure. There we also found the entanglement entropy was related to the thermal calculation by a factor of two.  Thus the single-sided black string also agrees with CFT, for both the half-line and the interval, and provides a precise holographic dual of the Cardy-Calabrese B-state computations.

\newsec{ A comment on the long time expansion of correlators}

In the previous sections we pointed out that the entanglement entropy had a piece linear in time
which was produced by a surface living in the interior at some particular interior-time $\rho = i\kappa_m$. Its extension in
the spacelike $t_I$-direction in the interior gives rise to the piece linear in time.
In this section we ask whether the same behavior is present for other observables.
Indeed,  we will  find a similar
behavior for  correlators of operators whose duals are heavy particles, strings or branes.

First we note that in the case of the BTZ black hole in $AdS_3$, the minimal length computation we did
for the entanglement entropy is formally identical to the one we would do for computing the
correlator of two operators, one on each boundary, in the regime where we use the classical geodesic
approximation in the bulk ($\Delta \gg 1 $). The fact that the action is linear in time
leads to the exponential in time behavior of the correlator $ \langle O_1(t_b)  O_2(t_b) \rangle
\sim e^{ - 2 t_b m R_{AdS_3} } $. This decay rate is simply the lowest quasinormal mode associated to
the corresponding bulk field.

In fact, in any dimension,
 in the large mass regime,
  the lowest quasinormal mode is related to a geodesic that
sits at a particular (possibly complex) value of $\rho$ in \btzc , and is extended along the
time direction. The value of $\rho$ is determined by extremizing the action for a geodesic
\eqn\acgeo{
  i S = - \int dt \sqrt{-g^2 + \dot \rho^2 }
  }
  after setting $\dot \rho =0$. Thus we
  extremize $\sqrt{-g^2 }$.  In general the extrema lie at complex values of $\rho$. For example for the black brane metric \gendi\
  we find that
  \eqn\metrx{
    \cosh (d \rho_c) = - (d-1) ~,~~~~  \sqrt{-g^2} = \pm e^{ \pm i { \pi \over d } } \tilde c_d
    }
    where $ \tilde c_d$ is a positive function of $d$ \foot{ $\tilde c_d ={ 2 \over d }
     \left( { d -2 \over 2 } \right)^{1 \over d } \sqrt{ d \over d -2 } $, which differs by a simple factor
     from eqn (5.3) in \FestucciaZX , due to a different choice of the temperature.
         In evaluating $\sqrt{ - g^2}$ we need
     to take a $d$th root, which we did following the prescription in \FestucciaZX .}.
      These values agree with the quasinormal frequencies
    discussed in great detail by Liu and Festuccia in \FestucciaZX . In this case we get a complex value
    of $\rho_c$ so we cannot say that the geodesic is in the interior. What we wanted to highlight here is
    the simple fact that the geodesic is extended along the $t$-direction.  The saddle points in \metrx\ are complex because the function $\sqrt{ - g^2}$ does not have a maximum in
 the interior of the black hole, it grows from the horizon to the singularity.

    If one considers particles with imaginary momentum, $p = i k m $, with real $k$, then for
    $ 0 < k < 1$ there are quasinormal modes which correspond to particles sitting in the interior
    at real values of the interior time coordinate $ \kappa$  ($\rho = i \kappa $). As $k \to 1$ these
    geodesics move closer to the horizon \FestucciaZX .

    Note that the computation of quasinormal modes can be done by solving the wave equation completely outside
    the horizon. Nevertheless the actual answer is well approximated by geodesics that lie inside.  This
    was explored in detail in \refs{\FidkowskiNF,\FestucciaPI}.
     In particular the solution of the problem
    outside dictates which geodesics should be included. On the other hand, in the entanglement entropy computation that
    we did above, we included only the real geodesic but no other possible complex minimal surfaces. Presumably a more
     detailed analysis of possible geodesics would show that the real ones are the dominant ones. This is a gap in our
      reasoning that we leave to the future. We have also ignored
    the possibility that the surfaces could hit the singularity.

We now consider expectation values of Wilson loops and higher dimension
surface defect operators.
 More precisely, we   imagine correlators of straight surface operators,
  extended along the spatial directions, and localized in the time direction. We
   consider two such operators, one in each of
    the two copies of the thermofield double at equal times $t_b$.
  Then to extract the large $t_b$ behavior,
  we need
 to consider a surface extended along the spatial directions and along the (spatial) $t_I$-direction of
 the interior. To find the interior position in the timelike direction we need to minimize
 \eqn\minqu{
  i S  = -   T   \sqrt{ - g^2(\rho) } \left[ h(\rho) \right]^{p}
    }
    where $p$ is the dimensionality of the defect operator, $p=1$ for Wilson loops, $p=0$ for ordinary local
    operators, etc.  We see that the computation of the entanglement entropy is similar to
    a computation of a defect operator correlator with $p=d-2$.
Minimizing \minqu\  we get
\eqn\resol{
 \cos d \kappa_c  =  1 - { d    \over (1+p) }  \ .
    }
    The solutions are real as long as $ p \geq {  d - 2 \over 2 } $ (for the equal sign we have surfaces that
    approach the singularity). Note that for $d=4$ and $p=1$ we get a surface that is very close to the singularity.

    The expectation value of the defect operator then goes like
    \eqn\deifoc{
    \langle W \rangle \sim e^{ - 2 t_b  ( -i S_c) }
    }
    where $S_c$ is the value of \minqu\ at \resol .
    A similar expression gives the decay of the expectation value of a defect operator for a black hole
    that forms from collapse, except that $ 2 t_b \to t_b$.

\newsec{ Tensor networks and time evolution }

In this section we suggest a tensor network representation for the wavefunction of the thermofield double or the
pure B-state which is inspired by the above computations.
A tensor network is a representation of the wavefunction for a discrete quantum system that captures the relevant subspace in the Hilbert space where
the wavefunction in question lives. This has been discussed both for massive theories and conformal field theories \VidalRev .
It incorporates efficiently the information about the renormalization group.

\ifig\MassiveOneD{  Tensor network representation of the wavefuction of a gapped two dimensional system.
 The vertical open lines represent each of the
spins of a one dimensional chain. The intersection points with the horizontal line represent tensors. The horizontal line
gives the pattern of index contractions of these tensors. For each value of the spin (the index of the vertical line), the
tensor has two indices, the two horizontal lines coming out of the vertex. These are contracted with the neighboring tensors.
   } {\epsfxsize3.5in\epsfbox{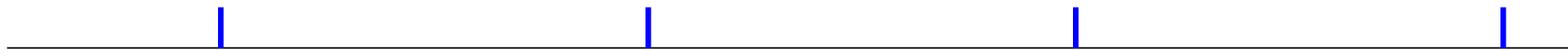}}

The simplest case is a spin system in one space dimension in a state with a mass gap.
The tensor network representation of the wavefunction (called a matrix product state  in this case \OstlundZZ ) has the form
\eqn\tensne{
\Psi( s_1 , \cdots , s_L ) = Tr[ T_{s_1} T_{s_2} \cdots T_{s_L} ] \ ,
 }
 where $ T_{s_i} = (T_{s_i})^k_l $ is a $D\times D$ matrix and $s_i = \pm 1 $ are the values of the spin at each site.
 See figure \MassiveOneD .
 This can be viewed as a particular ans\"{a}tz for the wavefunction that parametrizes the subspace of the Hilbert space where we expect the groundstate wavefunction to live. It
 is a space of dimension $ 2 L D^2$  which is smaller than the $2^L$-dimensional total Hilbert space of the system.
 We need $\log D $  comparable to the
 entanglement entropy of half the system in order to get a reasonable answer. This allows for a drastic reduction in the size of the relevant Hilbert space and has proved very useful for the efficient numerical computation of the groundstate wavefunction in 1d gapped systems \WhiteZZ\ (see \SchollRev\ for a recent review).

\ifig\MassiveTwoD{   Pattern of index contractions for the tensor network representation for the ground state of a
massive theory in two spatial dimensions.
   } {\epsfxsize3.5in\epsfbox{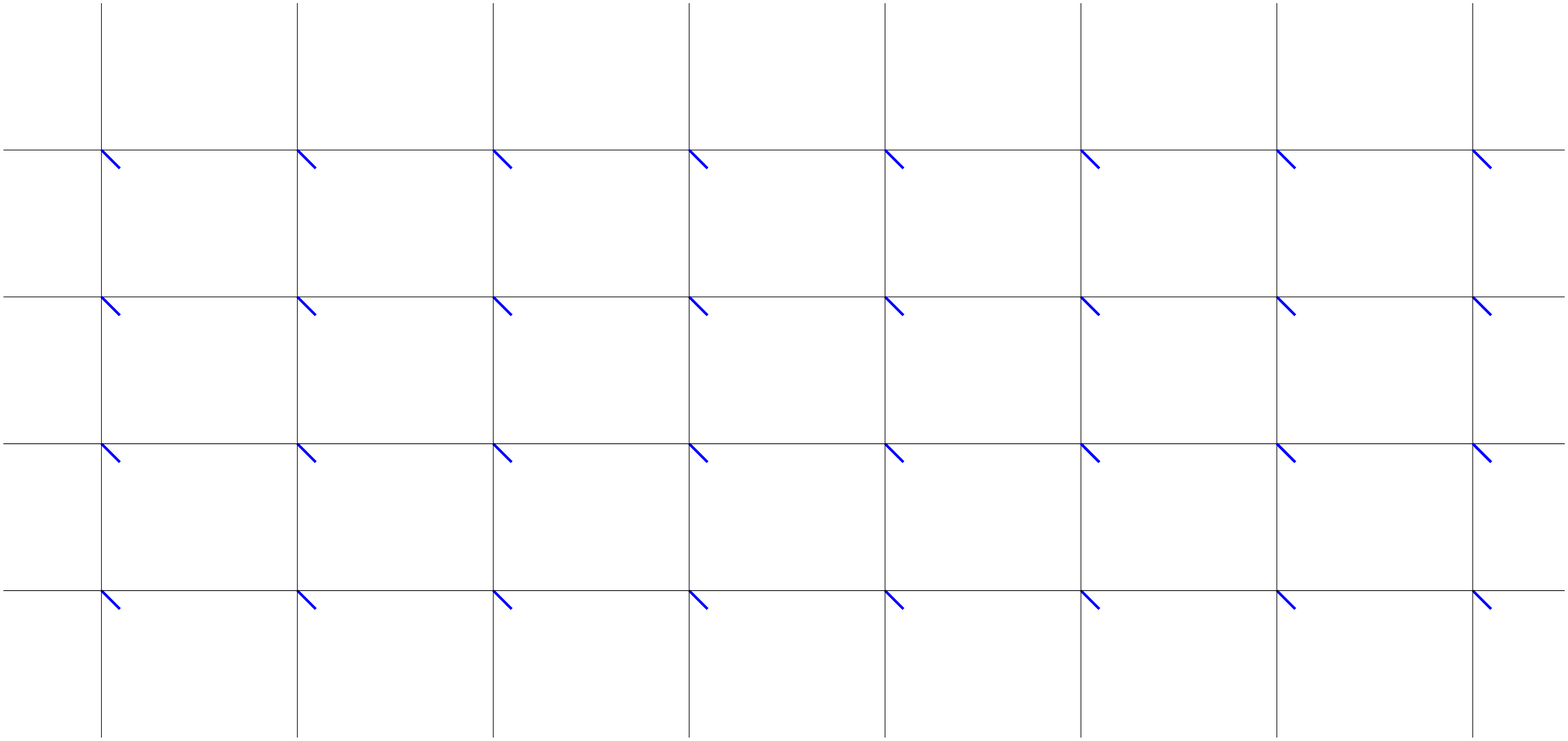}}

 One can write similar ans\"atze for  gapped  systems in $d$ spatial dimensions.
  In that case the tensors $T$ will have $2^d$ indices and
 their contraction gives a structure similar to a $d$ dimensional lattice, see \MassiveTwoD ,  \cirac .

\ifig\Mera{   Tensor network representation for a conformal theory in the IR. At each intersection point we have a
tensor with five indices and the pattern of index contractions that is indicated by the diagram.
   } {\epsfxsize3.5in\epsfbox{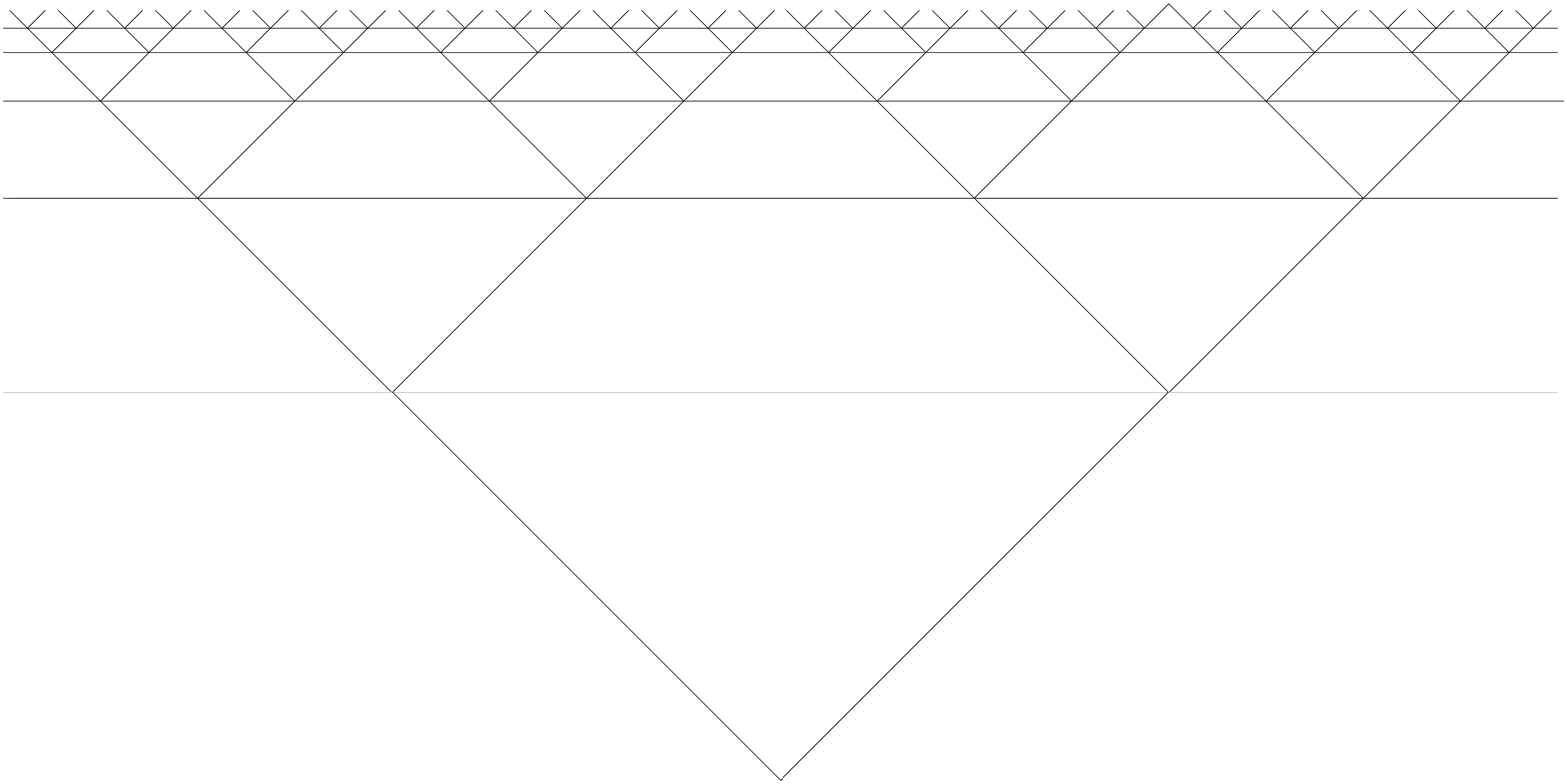}}

 For systems that are conformal in the IR there is also a convenient representation which is called the
 MERA network \VidalMera. It is a particular tensor network with contractions as indicated in \Mera  .
 For more details see \VidalRev. This is a representation of the wavefunction which incorporates the renormalization
 group. More precisely, it can realize explicitly a discrete subgroup of the renormalization group. As pointed out
 in \SwingleBG, we can think of these tensors as the wavefunctions of regions with size of order the $AdS$ radius\foot{
 This represents locality well for distances large compared to the $AdS$ radius, but not for smaller distances. After all,
 such small-scale locality is not valid for an arbitrary CFT.}.
 These tensor networks nicely capture some features of the entanglement entropy, as discussed in detail in \refs{\SwingleBG,\VidalHolo,\SwingleWQ,\NozakiZJ}. For the 1d chain (\MassiveOneD), the entanglement entropy of half the system is bounded above by $\log D$, where $D$ is the dimension of the link that must be cut to separate the system into two parts.  In a more elaborate tensor network, if we must cut $\ell$ links to separate subsystem $A$ from the rest of the tensor network, then
\eqn\salse{
 S_A \lesssim \ell \log D \ .
}
The MERA network is a discretized version of hyperbolic space, and we can separate system $A$ by cutting the links on a minimal-size surface extending into the tensor network.  Therefore, if the tensors at each site are sufficiently random for the bound \salse\ to be saturated, then this resembles the Ryu-Takayanagi formula for holographic entanglement entropy
\SwingleBG .

\ifig\ThermalMera{   Pattern of index contractions for the tensor network representation for the thermofield double
 of a scale invariant theory. The scale invariance is broken by the temperature. At scales of order the temperature we
 have a pattern similar to the one we had for the massive theory in \MassiveOneD .
   } {\epsfxsize3.5in\epsfbox{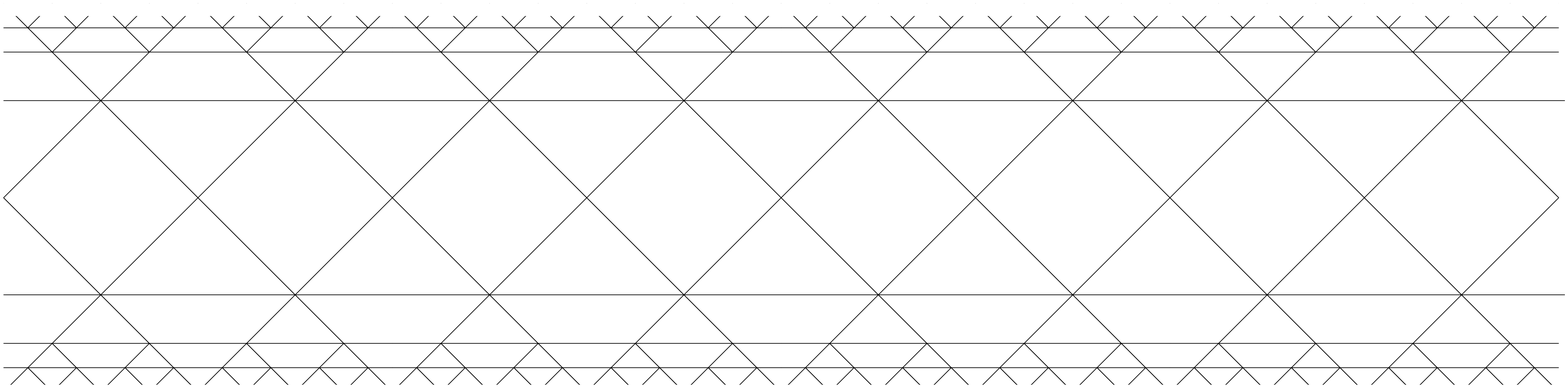}}

 \ifig\ThermalMeraEvolved{   Pattern of index contractions for the tensor network representation for the thermofield double
 of a scale invariant theory after time evolution. As we evolve we add of order $t/\beta$ intermediate layers. This picture
 is suggested by the geometry of the interior and general properties of time evolution. The red dotted line cuts the
 system into   halves, and the entanglement entropy is bounded by the number of black lines that it crosses.
   } {\epsfxsize3.5in\epsfbox{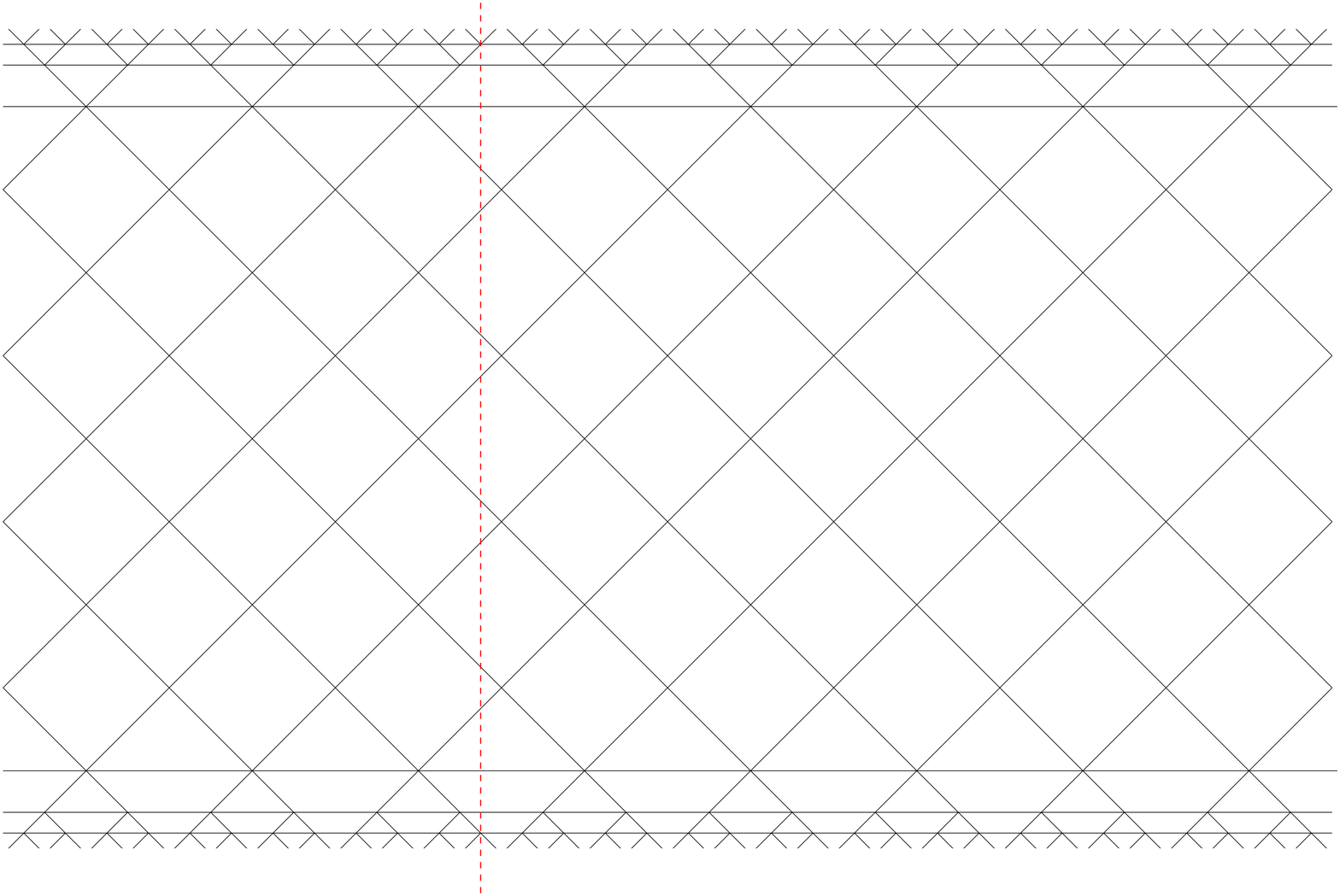}}

 The thermofield double or the $B$-state are gapped states with a scale invariant UV region. Thus the tensor
 network representation of those states are expected to be as in \ThermalMera . We essentially have the MERA ans\"{a}tz up
 to the scale of the temperature and then we have a gapped state. This is the tensor network representation of
 the state at $t=0$. We can now ask, what is the representation of the state at larger times $t$.
 Our entanglement computation leads to a natural proposal. We need to use a tensor network that grows as $t$ grows.
 We simply add more layers in the middle region. We expect the number of these layers to be of order $t/\beta$.
 The resulting tensor network in this middle region looks similar to the one we would obtain for a gapped system in
 $d$ spatial dimensions. The structure of the network is similar to the structure of the nice slices. The gapped region
 is the interior of the horizon. The fact that it is gapped is associated to the fact that all correlators decay
 exponentially in time. This representation of the wavefunction of such systems does not appear to have been used in
 the tensor network literature, but the black hole suggests that it should be useful.

 At some level the representation in terms of tensors as in   \ThermalMeraEvolved\ is obvious, since we can view each additional row
 that we add, as we time evolve, as the action of the Hamiltonian on the block spin variables at scale $\beta$. The
 quick thermalization is implying that the resulting state is similar to that of a gapped system,
 at least for simple local  operators
 (the full theory is still unitary).

 We know that for long times we cannot have a perfect decay for all correlators, since that would imply information loss.

 Notice that this tensor network idea is not good enough, so far, to resolve distances smaller than the $AdS$ radius. The size of the time-like $\rho$-direction in the interior is of the order of the $AdS$ size and it is not visible in this ans\"{a}tz for the wavefunction.
 On the other hand, the spacelike $t$-direction in the interior has a  very large size, and is visible.  Unfortunately since we cannot resolve the timelike $\rho$-direction, we cannot say whether we are talking about the interior or the region near but outside the horizon.

 \newsec{ Discussion }

 Note the peculiar nature of the entanglement at $t=0$. At $t=0$ the entanglement is
 relatively short range in the non-compact spatial directions. Degrees of freedom are entangled within a distance of
 order $\beta $. This is true both for the two sided and the one sided configurations. For this reason the
 entanglement entropy obeys an area law, similar to the one observed in theories with a mass gap.
 In fact, it is well known that thermal field theories behave as systems with a mass gap, when we consider equal time
 observables.  

 \ifig\particlesdouble{
Quasiparticle picture of growing entanglement in the thermal double CFT. Particles connected by a dashed line in the Euclidean region are entangled. One particle of the entangled pair is in each copy.
 At time $=0$ these correlations are localized around $x \sim \beta/2$. As time passes the particles move in the two copies of the CFT. Particles that contribute to the entanglement are those where one member of the pair is in region $A$ and the other is outside. For example the pair with the dashed blue line does not contributed because both particles are in region $A$.
   } {\epsfxsize 10cm
   \epsfbox{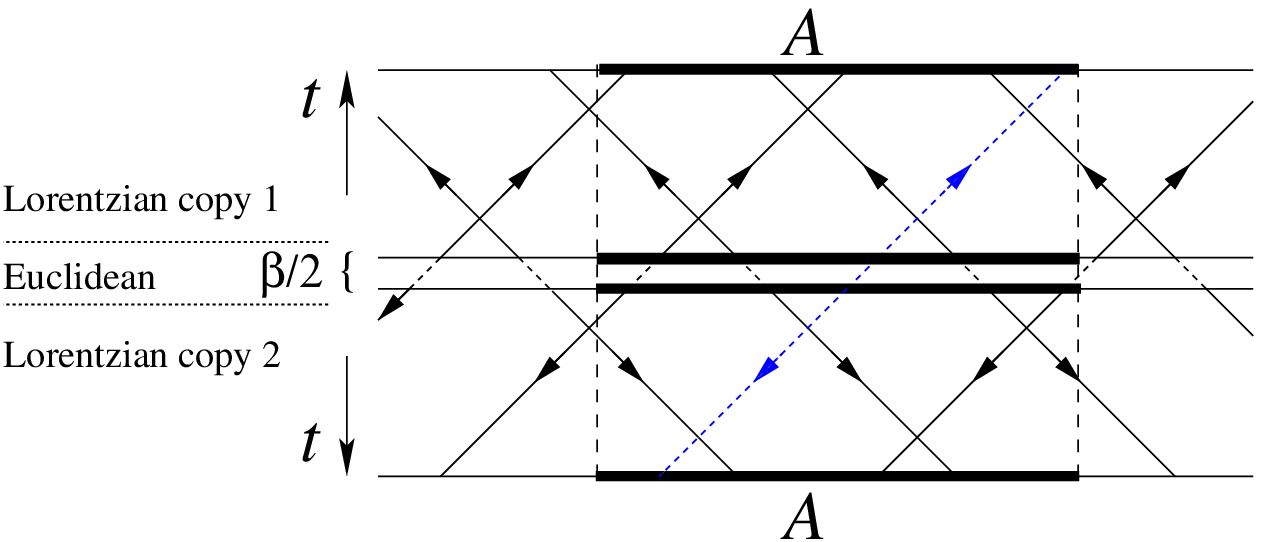}
   }

\ifig\particlessingle{
Quasiparticle picture of growing entanglement in the pure B-state. Euclidean time evolution creates entangled particles separated by $x \sim \beta/2$, heading in opposite directions. As time evolves one member of the pair can leave the region
$A$. This will increase the entanglement entropy. Here the pair with the dotted blue line does not contribute to the entanglement, but the rest do contribute.
   } {\epsfxsize 10cm
   \epsfbox{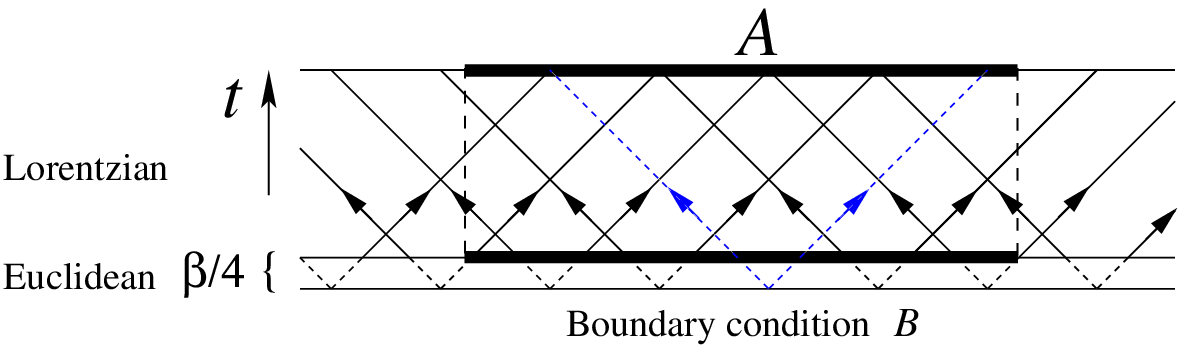}
   }

 As time evolves this short range entanglement spreads out along the spatial directions. This
 is what gives rise to the linear growth of the entanglement entropy in the boundary theory.
 In fact, we have the picture   in \particlesdouble\ and \particlessingle , discussed in \CalabreseIN  .
 This picture is rather obvious from the boundary theory point of view.
 It is  interesting that in the gravity description the same linear growth comes from the linear growth
 of the spacelike $t$-direction in the
  interior region of the black brane geometry. This linear growth in time and  the fact that
 it comes from the interior region was pointed out in previous work
 \refs{\HubenyXT,\AbajoArrastiaYT,\AparicioZY,\AlbashMV,\BalasubramanianCE,\BalasubramanianUR,%
\AsplundCQ,\BasuFT,\BasuGG,\BuchelLLA,\BalasubramanianAT,\AllaisYS}.   We hope to have clarified and isolated the
 origin of this linear behavior as arising from a very special spatial slice in the interior. It is the slice that
 maximizes the area for a codimension two surface.
It is also noteworthy that \particlesdouble\ and \ThermalMeraEvolved\ look  similar. In fact if we
look at \particlesdouble\ and we view the lines as representing index contractions and the intersections as representing
tensors, then it looks rather similar to \ThermalMeraEvolved .

 In the boundary theory, when we disturb the fluid with the insertion of a local operator, we create a local disturbance
 that is then carried to longer and longer distances.  The above results for entanglement entropy suggest that
 when the excitations have reached a size bigger than of order $\beta$, then we should think of them as living in the
 interior. As time progresses these excitations live further and further away in the interior, as measured along a
 spacelike ``nice'' slice. Since its information is spread out over a long distance it seems reasonable that it would
 be hard to extract via simple local operators.
 Of course, this is not explaining why this excitation cannot get out of the black hole. To understand bulk causality, we
 need to understand bulk locality at distances less than the $AdS$ radius.

 Throughout this paper we have assumed that we have a non-compact direction. If we have the QFT on a space of size
 $L$, then we can follow the growth of the entanglement entropy in this fashion up to a time of order $L$. On the
 other hand, the interior seems to continue making sense for times which are much longer.  It is probable that
 the ``long string'' phenomenon \MaldacenaDS\  is relevant for enabling the spread of the entanglement.
 In other words, it is likely that the interior is associated to the spread of entanglement, but that such a spread is
 occurring among the matrix degrees of freedom rather than along the physical spatial dimensions.
 An extreme example could be the black hole in the gravity dual of the D0 brane quantum mechanics (or any other
 gravity dual of a quantum mechanics theory). In this case, there are no extra spatial directions and the holographic
  direction,
 as well as the interior of the black hole, has to come solely from the structure of the matrix degrees of freedom.

Recently it was suggested that  black holes which are
maximally entangled with radiation do not have smooth horizons \AlmheiriRT.
The eternal black hole is an example of a black hole such that its microstates,
given by the CFT on one side, are maximally entangled with a second system, the
second copy of the CFT. The two CFT's do not talk to each other and we can view the
second CFT as living in the memory of the quantum computer that has processed
the outgoing radiation\foot{It has been claimed that it is impossible to construct such a computer \HarlowTF.}.
In this particular case, we get that the horizon is indeed smooth.
Of course, this state is very special. One can wonder what happens
if we applied a generic unitary transformation to the
Hilbert space of the second side. Could that create a firewall?
One unitary transformation that we can apply is simply time evolution.
We see that time evolution does not tend to destroy the horizon. On the contrary,
it is  making the interior region even bigger, in the sense described above.
Namely, time evolution gives rise to the spacelike $t$-direction in the interior region, labeled
$In$ in figure \Penrose . However, this does not address the general case.

\bigskip

\noindent{\bf Acknowledgments}

\noindent
It is a pleasure to thank  G. Festuccia, D. Harlow, I. Klebanov, H. Liu and  S. Sachdev  for useful discussions. This work was supported in part by U.S.~Department of Energy grant DE-FG02-90ER40542. T.H. also acknowledges support from the Corning Glass Works Foundation Fellowship Fund.

 \appendix{A}{ Change of coordinates for the black brane }

 The usual coordinates for a black brane for $AdS_{d+1}$ have the form
 \eqn\usuc{
 ds^2 = { 1 \over z^2 } \left[ - ( 1-z^d) dt^2 + { dz ^2 \over (1-z^d)} +   d x^2  \right] \ .
 }
 Here the temperature is not $1/(2\pi)$, but we will fix that later.
 We now write
 \eqn\eqnsfi{ \eqalign{
 d\rho = &  { d z \over z \sqrt{(1 -z^d)} } ~,~~~~~~~~  \tanh { d \rho \over 2 } = \sqrt{ 1 - z^d } \ .
 }}
 Then $h$ and $g$ are as given in \gendi , where we rescaled $t$ and $x$ to introduce the factor of
 ${ 2 \over d }$ to fix the temperature to $1/(2 \pi )$.

\appendix{B}{ Coordinates of the BTZ string}
In this appendix we relate the various coordinate patches of the BTZ string to Poincar\'{e} coordinates via the embedding space
\eqn\embedi{
-Y_{-1}^2- Y_0^2 + Y_1^2 + Y_2^2 = -1 \ .
}
The Poincar\'{e} coordinates \btzpoin\ are
\eqn\embedpoin{
Y_{-1}={1\over 2z}[1+(z^2-x_0^2+x_1^2)],\quad Y_2 = {1\over 2z}[1- (z^2-x_0^2+x_1^2)], \quad Y_0 = {x_0\over z}, \quad Y_1 = {x_1\over z} \ .
}
The exterior coordinates \btzc\ are
\eqn\embedext{
Y_{-1} = \cosh x \cosh\rho , \quad Y_2 = -\sinh x \cosh\rho , \quad Y_0 = \sinh t \sinh\rho , \quad Y_1 = \cosh t \sinh\rho \ ,
}
which covers $-Y_0^2 + Y_1^2 \geq 0, -Y_{-1}^2 + Y_2^2 \leq -1$. The interior coordinates \btzin\ are
\eqn\embedint{
Y_{-1}=\cosh x \cos\alpha,\quad Y_2=-\sinh x \cos\alpha,\quad Y_0 = \cosh\tilde t \sin\alpha,\quad Y_1 = \sinh\tilde t\sin\alpha \ .
}
The interior covers $-Y_0^2 + Y_1^2 \leq 0,-Y_{-1}^2+Y_2^2\leq0$, or $-1 \leq -Y_0^2 + Y_1^2 \leq 0$.  Finally the future region \btzfuture\ is related to the embedding space by
\eqn\embedfuture{
Y_{-1}=\sinh\tilde x \sinh\tilde\rho,\quad Y_2 = -\cosh\tilde x \sinh\tilde\rho,\quad Y_0 =\cosh\tilde t\cosh\tilde\rho,\quad Y_1 = \sinh\tilde t\cosh\tilde\rho \ .
}
This region covers $-Y_0^2 + Y_1^2 \leq -1, -Y_{-1}^2 + Y_2^2 \geq 0$.

\listrefs

\bye